\documentclass[aps,prd,twocolumn,superscriptaddress,showpacs,showkeys]{revtex4}

\usepackage{epsfig,epsf}
\usepackage{amsmath}
\usepackage{amsthm}
\usepackage{amsfonts}
\usepackage{amssymb}
\usepackage{dsfont}

\usepackage{multirow}

\usepackage{slashed}

\usepackage[active]{srcltx}
\usepackage{psfrag}

\newcommand{\be}{\begin{equation}}
\newcommand{\ee}{\end{equation}}
\newcommand{\ba}{\begin{eqnarray}}
\newcommand{\ea}{\end{eqnarray}}
\newcommand{\baa}{\begin{eqnarray*}}
\newcommand{\btab}{\begin{tabular}}
\newcommand{\etab}{\end{tabular}}
\newcommand{\eaa}{\end{eqnarray*}}

\def\inbar{\,\vrule height1.5ex width.4pt depth0pt}
\def\IC{\relax\hbox{$\inbar\kern-.3em{\rm C}$}}
\def\IZ{\relax{\hbox{\cmss Z\kern-.4em Z}}}
\def\IR{{\hbox{{\rm I}\kern-.2em\hbox{\rm R}}}}

\def\IP{{\hbox{{\rm I}\kern-.2em\hbox{\rm P}}}}
\def\II{\hbox{{1}\kern-.25em\hbox{l}}}

\begin{document}

\title{Light Cone Sum Rules for the $\pi^0\gamma^*\gamma$ Form Factor Revisited}


\date{\today}

\author{S.S.~Agaev}
\affiliation{Institut f\"ur Theoretische Physik, Universit\"at
   Regensburg, D-93040 Regensburg, Germany}
\affiliation{Institute for Physical Problems, Baku State University,
 Az--1148 Baku, Azerbaijan}
\author{V.M.~Braun}
\affiliation{Institut f\"ur Theoretische Physik, Universit\"at
   Regensburg, D-93040 Regensburg, Germany}
\author{N.~Offen}
\affiliation{Institut f\"ur Theoretische Physik, Universit\"at
   Regensburg, D-93040 Regensburg, Germany}
\author{F.A.~Porkert}
\affiliation{Institut f\"ur Theoretische Physik, Universit\"at
   Regensburg, D-93040 Regensburg, Germany}

\begin{abstract}
 We provide a theoretical update of the calculations of the
$\pi^0\gamma^*\gamma$ form factor in the LCSR framework,
including up to six polynomials in the conformal expansion
of the pion distribution amplitude and taking into account
twist-six corrections related to the photon emission at large distances.
The results are compared with the calculations
of the $B\to\pi \ell\nu$ decay and pion electromagnetic form factors
in the same framework. Our conclusion is that the recent BaBar
measurements of the $\pi^0\gamma^*\gamma$
form factor at large momentum transfers
\cite{BABAR} are consistent with QCD,
although they do suggest that the pion DA may have more structure than
usually assumed.
\end{abstract}

\pacs{12.38.Bx, 13.88.+e, 12.39.St}

\keywords{exclusive processes; form factor; sum rules}

\maketitle


%
\section{Introduction}
%

Despite a solid theory background~\cite{Chernyak:1977as,Radyushkin:1977gp,Lepage:1979zb},
phenomenological success of QCD in exclusive reactions has been rather modest.
A problem is that, since the quarks carry only some fractions
of hadron momenta, virtualities of the internal lines
of the hard subprocess appear to be essentially smaller than $Q^2$, the nominal
momentum transfer to the hadron.
As a result, at accessible $Q^2$, the  bulk part of the hard QCD contribution
comes from the regions where the ``hard'' virtualities are
much smaller than  the  typical hadronic scale of 1\,GeV$^2$
\cite{Efremov:1980mb,Isgur:1988iw,Radyushkin:1990te}.
According to the factorization principle, contributions from such regions
have to be subtracted from the hard coefficient function and included separately
as additive ``soft'' or ``end-point'' nonperturbative contributions.
The standard power counting suggests that ``soft'' terms are
suppressed by extra powers of $1/Q^2$.
However, they do not involve small coefficients $\sim \alpha_s(Q)/\pi$ which are
endemic for factorizable QCD contributions based on hard gluon exchanges.
As the result, the onset of the perturbative regime may be postponed to very
large momentum transfers.

The pion transition form factor ${\gamma^*\gamma^{(*)} \to \pi^0}$
with at least one virtual photon plays a very special r\^ole as it
is the simplest hard exclusive process where the above mentioned
difficulties are absent or, at least, moderated. There is only one
hadron (pion) involved, and the large $Q^2$ behavior of this form
factor is determined  \cite{Lepage:1979zb} by the operator product
expansion (OPE) of the product of two electromagnetic currents
near the light cone which is very well studied in the context of
inclusive deep-inelastic scattering (DIS). In this case the leading
contribution to the hard coefficient function is of
order one (i.e. not suppressed) as no gluon exchanges are
involved, and at the same time ``soft'' (end-point) contributions
either do not exist --- for the case of two virtual photons --- or
are likely to be suppressed, if one photon is real. These features
make the pion transition form factor an ideal place to test the
QCD factorization approach and determine the pion distribution
amplitude (DA) which can then be used to describe other exclusive
hard reactions. This task is as important as ever, the most
high-profile application being at present to exclusive weak
$B$-decays, $B\to \pi \ell \nu_\ell$, $B\to \pi\pi$ etc. which are
the main source of precision information on quark flavor mixing
parameters in the Standard Model. These are the aim of an
extensive experimental study: It addresses the question whether
there is New Physics in flavor-changing processes and where it
manifests itself.

Whereas the case of two virtual photons offers crucial
simplifications for the theory, the transition form factors with
one real and one virtual photon are much easier to study
experimentally. They can be measured for space-like momentum
transfers in the process $e^+ e^- \to e^+ e^- \pi^0,\eta,\ldots$
and in $e^+ e^- \to \gamma \pi^0,\eta,\ldots$ for time-like ones.
Till 1995 only the CELLO data \cite{Behrend:1990sr} were available
at relatively low, space-like momentum transfer: $Q^2 <
2.5$~GeV$^2$ for $\pi^0(\eta)\gamma^*\gamma$ and somewhat higher
for $\eta'\gamma^*\gamma$. The covered region was extended to $Q^2
\sim 9-15$~GeV$^2$ by CLEO collaboration \cite{Gronberg:1997fj}
which allowed, for the first time, a quantitative comparison with
the perturbative QCD. The CLEO data appeared to be consistent with
the predicted scaling behavior $\sim 1/Q^2$ setting in for
momentum transfers of the order of a few GeV$^2$ and also
suggested that the pion DA is somewhat broader compared to its
asymptotic shape at large scales, which was, again, expected based
on the corresponding studies using QCD sum rules. More recently,
BaBar reported \cite{Aubert:2006cy} a measurement of the time-like
transition form factors  $\eta\gamma^*\gamma$ and
$\eta'\gamma^*\gamma$ at very large $q^2 = 112$~GeV$^2$. No
significant tension with the theory was observed, although
predictions in the time-like region are generally more difficult.

The situation changed in 2009 when the BaBar collaboration presented \cite{BABAR}
the measurement of the $\pi^0\gamma^*\gamma$ form factor up to photon
virtualities of the order of 40~GeV$^2$. These new data
created considerable excitement in the theory community as they
do not show the expected scaling behavior. The most popular explanation
so far has been \cite{Radyushkin:2009zg,Polyakov:2009je} that the pion DA
has an unexpected ``flat'' shape and does not vanish at the end points.
This, on one hand, triggered speculations on the breakdown of QCD factorization
\cite{Dorokhov:2010bz} and, on the other hand, claims that the BaBar data are in contradiction
with light-cone sum rules (LCSRs) and with the common wisdom  on the
lowest moments of the pion DA \cite{Mikhailov:2009kf,Mikhailov:2010aq,Mikhailov:2010ud}.
The aim of this work is to reexamine these claims by making an updated analysis
of the $\pi^0\gamma^*\gamma$ form factor within the LCSR approach
including up to six polynomials in the conformal expansion
of the pion distribution amplitude and taking into account
twist-six corrections. The photon emission at large distances is
discussed in detail. The results are compared with the calculations
of pion electromagnetic form factor and $B\to\pi \ell\nu$ decay
in the same framework. Our conclusion is that the recent BaBar
measurements \cite{BABAR} are consistent with QCD and with
the bulk of the available information on the pion distribution amplitude.
In particular we argue that the ``flat'' DA \cite{Radyushkin:2009zg,Polyakov:2009je}
is not warranted and in fact no conclusion on the end--point behavior
of the DA can be inferred on the basis of the existing experimental data.

The presentation is organized as follows. Sect.~2 contains a
concise review of the QCD (collinear) factorization approach to
the $\pi^0\gamma^*\gamma$ form factor and the existing information
on the pion DA. The LCSR approach is motivated and explained in
detail in Sect.~3. In this work we go beyond the existing analysis
\cite{Khodjamirian:1997tk,Schmedding:1999ap,Bakulev:2001pa,Bakulev:2002uc,Bakulev:2003cs,Agaev:2005rc,Mikhailov:2009kf}
in two aspects. First, we calculate a new, twist-six contribution
to LCSRs which proves to be sizeable. This contribution is related
to photon emission from large distances for which we also derive
the leading-order perturbative expression. Second, we extend the
existing formalism to allow for the contributions of higher-order
Gegenbauer polynomials, which allows one to consider DAs of
arbitrary shape and also address the question of convergence of
the Gegenbauer expansion which generated some confusion. The
second task was already addressed in Ref.~\cite{Mikhailov:2009kf},
but our expressions do not agree, unfortunately. Sect.~4 contains
the numerical study of the LCSRs to the NLO accuracy. We consider
various uncertainties of the method in some detail and provide
error estimates for our predictions. The final Sect.~5 is reserved
for a summary and conclusions.

%
\section{QCD factorization and pion DA}
%

%
\begin{figure*}[t]
\begin{center}
\includegraphics[width=.60\textwidth,
clip=true]{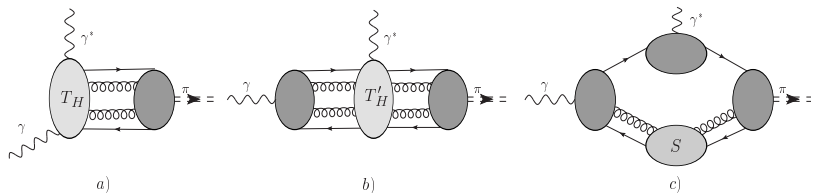}
\end{center}
\caption{Schematic structure of the QCD factorization for the $F_{\gamma^*\gamma \to \pi^0}(Q^2)$
factor factor.}
\label{fig:flow}
\end{figure*}

The form factor $F_{\gamma^*\gamma^* \to \pi^0}(q_1^2,q_2^2)$
describing the pion transition in two (in general virtual) photons
can be defined by the matrix
element of the product of two electromagnetic currents
\begin{eqnarray}
\lefteqn{\hspace*{-0.5cm}
 \int d^4 y e^{iq_1y}\langle\pi^0(p)|T\{j^{\rm em}_\mu(y)j^{\rm em}_\nu(0)\}|0\rangle =}
\nonumber\\
&&\hspace*{1.5cm}=\,  i e^2 \varepsilon_{\mu\nu\alpha\beta} q_1^\alpha q_2^\beta F_{\gamma^*\gamma^* \to \pi^0}(q_1^2,q_2^2)
\label{eq:Fgamma}
\end{eqnarray}
where
$$j^{\rm em}_\mu(y) = e_u\bar u(y)\gamma_\mu u(y) +  e_d\bar d(y)\gamma_\mu d(y)+\ldots,$$
$p$ is the pion momentum and $q_2 = q_1 +p$. We will consider the space-like form factor,
in which case the both photon virtualities are negative.
The experimentally relevant situation is when one virtuality is large
and the second one small (or zero). For definiteness we take
\begin{equation}
 q_1^2 = -Q^2\,,\qquad q_2^2 = -q^2\,,
\end{equation}
assuming that $q^2 \ll Q^2$. Most of the equations are written for $q^2=0$
and we use a shorthand notation
$F_{\gamma^*\gamma \to \pi^0}(Q^2)\equiv F_{\gamma^*\gamma^* \to \pi^0}(q_1^2 = -Q^2,q^2=0)$.

In general, a power-like falloff of the form factor $F_{\gamma^*\gamma \to \pi^0}(Q^2)$
in the large-$Q^2$ limit can be generated by the three different possibilities
of the large-momentum flow as indicated schematically in
Fig.~\ref{fig:flow}~\cite{Musatov:1997pu}.
The first possibility, Fig.~\ref{fig:flow}a, corresponds to the hard subgraph that
includes both photon vertices.
This is the dominant contribution that starts at order $\sim 1/Q^2$. For zero (or small,
$q^2 \leq \Lambda^2_{\rm QCD}$) virtuality of the second photon there exists another
possibility shown in Fig.~\ref{fig:flow}b:  In this case the low-virtual
photon is emitted at large distances and the large momentum flows through a subgraph
corresponding to hard gluon exchange between the quarks. The power counting for this
contribution shows that it is at most ${\mathcal O}(1/Q^4)$, i.e. subleading compared to the
first regime. Finally, the third possible regime shown in  Fig.~\ref{fig:flow}c corresponds
to the Feynman mechanism, i.e. the possibility that the quark interacting with the
hard photon carries almost all the momentum whereas the quark spectator is soft.
This contribution can intuitively be thought of as an overlap of nonperturbative wave functions
describing the initial (photon) and final (pion) states. In perturbation theory,
this contribution also scales as ${\mathcal O}(1/Q^4)$ in the large-$Q^2$ limit.

The contribution in  Fig.~\ref{fig:flow}a by construction involves a time-ordered
product of two electromagnetic currents at small light-cone separations. Hence it
can be studied using Wilson operator product expansion. The leading contribution
 ${\mathcal O}(1/Q^2)$ to the form factor corresponds to the contribution of the leading
twist-two operators and can be written in the factorized form
\begin{equation}
  F_{\gamma^*\gamma \to \pi^0}(Q^2) = \frac{\sqrt{2}f_\pi}{3}\!\int_{0}^{1}\! dx\, T_H(x,Q^2,\mu,\alpha_s(\mu))\phi_\pi(x,\mu)\,,
\label{eq:lt}
\end{equation}
where $\phi_\pi(x,\mu)$, the pion distribution amplitude  at the scale $\mu$,
is defined by the matrix
element of the nonlocal quark-antiquark operator stretched along the light-like direction
$n_\mu$, $n^2=0$:
\begin{eqnarray}
\lefteqn{\hspace*{-1.2cm}
 \langle 0| \bar q(0)[0,\alpha n ]\not\! n \gamma_5  q(\alpha n)|\pi^+(p)\rangle =}
\nonumber\\
&&\hspace*{0.5cm}=\, i f_\pi p\cdot n\int_0^1 dx\, e^{-ix \alpha p\cdot n} \, \phi_\pi(x,\mu)\,.
\label{eq:defDA}
\end{eqnarray}
Here and below
$\bar q\,\slashed{n} \gamma_5  q =
(1/\sqrt{2})[\bar u\,\slashed{n} \gamma_5  u -\bar d\,\slashed{n} \gamma_5  d]$.
To this accuracy (leading twist) all gluon attachments to the hard subgraph
(cf.~Fig.~\ref{fig:flow}a) can be absorbed in the path-ordered gauge link (Wilson line):
\begin{eqnarray}
[0,\alpha n ] & = & ~\mbox{\rm Pexp}\left\{ -ig \int_{0}^{\alpha} du \, n^\mu A_\mu(u n)\right\}.
\label{eq:8}
\end{eqnarray}
The normalization is such that
\begin{equation}
 \langle 0| \bar q(0)\gamma_\nu\gamma_5
  q(0)|\pi^0(p)\rangle \,=\,
 i f_\pi p_\nu\,, \quad \int_0^1 dx\, \phi_\pi(x,\mu) \,=\,1\,,
\end{equation}
where $f_\pi \simeq 131$~MeV is the usual pion decay constant.

The coefficient function in (\ref{eq:lt}) is known in the
$\overline{MS}$ scheme to the next-to-leading order (NLO) in the
strong coupling \cite{delAguila:1981nk,Braaten:1982yp,Kadantseva:1985kb}.
Taking into account
the symmetry of the pion DA $\phi_\pi(x)=\phi_\pi(1-x)$ it can be written as
\begin{eqnarray}
  T_H^{\rm NLO} &=& \frac{1}{x Q^2}
\Big\{1+ C_F\frac{\alpha_s(\mu)}{2\pi}
\Big[\frac12 \ln^2 x -\frac{x\ln x}{2(1-x)}
\nonumber\\
&&{} -\frac92 + \left(\frac32+\ln x \right)\ln\frac{Q^2}{\mu^2}
\Big]\Big\}.
\label{eq:NLOcf}
\end{eqnarray}

Symmetry properties of the renormalization-group (RG) equation which governs the scale
dependence of the pion DA \cite{Radyushkin:1977gp,Lepage:1979zb} suggest the series expansion
of the DA in Gegenbauer polynomials
\begin{eqnarray}
 \phi_{\pi}(x,\mu) &=& \sum_{n=0}^\infty a_n(\mu)\, \varphi_n(x)\,,
\nonumber\\
 \varphi_n(x)&=&6x(1-x)C_n^{3/2}(2x-1)\,.
\label{eq:C32}
\end{eqnarray}
The first coefficient $a_0(\mu) = 1$ is fixed by the normalization condition whereas
the remaining ones, $a_n(\mu_0)$ for $n=2,4,\ldots$, have to be determined by some
nonperturbative method (or taken from experiment).

To leading order (LO) the Gegenbauer coefficients are renormalized multiplicatively
whereas to the NLO accuracy the mixing pattern becomes more complicated.
One obtains
\cite{Dittes:1983dy,Sarmadi:1982yg,Katz:1984gf,Mikhailov:1984ii,Mueller:1993hg,Mueller:1994cn}
\begin{eqnarray}
 a^{\rm NLO}_n(\mu) &=& a_n(\mu_0)\,  E^{\rm NLO}_n(\mu,\mu_0)
\nonumber\\
&&\hspace*{-1.2cm}{}+
 \frac{\alpha_s(\mu)}{4\pi}\sum_{k=0}^{n-2} a_k(\mu_0)\,  E^{\rm LO}_k(\mu,\mu_0)\,
  d_{n}^{k}(\mu,\mu_0)\,,
\label{eq:NLOevolution}
\end{eqnarray}
Explicit expressions for the RG factors $E^{\rm
(N)LO}_n(\mu,\mu_0)$ and the off-diagonal mixing coefficients
$d_{n}^{k}(\mu,\mu_0)$ in the $\overline{\rm MS}$ scheme are
collected in Appendix~A.

The NNLO calculations of the transition pion form factor exist in
the so-called conformal scheme
$\overline{\rm CS}$ \cite{Melic:2002ij,Braun:2007wv} but they cannot be converted
to $\overline{\rm MS}$ lacking the full NNLO result for the trace
anomaly term, which is so far not available.
An extensive discussion of scheme dependence can be found in
Refs.~\cite{Mueller:1993hg,Mueller:1994cn,Melic:2002ij}.

The existing information on the pion DA comes from QCD sum rules, lattice calculations
and light-cone sum rules. The first nontrivial Gegenbauer coefficient $a_2$ is related
to the second moment of the DA
\begin{equation}
 \langle \xi^2 \rangle \equiv \int_0^1 dx\, (2x-1)^2 \phi_\pi(x)\,,
\qquad a_2 = \frac{7}{12}\big(5\langle \xi^2 \rangle -1\big)
\label{eq:a2pi}
\end{equation}
and can be evaluated as a matrix element of the local operator with two derivatives
between vacuum and the pion state. There exists overwhelming evidence that this
coefficient is positive, meaning that the pion DA is broader than its asymptotic
expression $\phi_\pi^{\rm as} = 6x(1-x)$, see Table~\ref{tab:a2pi}.

\begin{table*}[t]
\renewcommand{\arraystretch}{1.2}
\begin{center}
\begin{tabular}{@{}l|l|l|l@{}} \hline
Method                      & $\mu=1$~GeV            & $\mu=2$~GeV      & Reference
                                                                                      \\ \hline
LO QCDSR, CZ model          & $ 0.56 $               & $0.39$        & \cite{Chernyak:1981zz,CZreport}
                                                                                      \\ \hline
QCDSR                       & $0.26^{+0.21}_{-0.09}$    &   $0.18^{+0.15}_{-0.06}$  & \cite{Khodjamirian:2004ga}  \\
QCDSR                       & $0.28\pm 0.08$            &   $0.19\pm 0.06$           & \cite{Ball:2006wn} \\
QCDSR, NLC                   & $0.19\pm 0.06$         &   $0.13 \pm 0.04$  & \cite{Mikhailov:1991pt,Bakulev:1998pf,Bakulev:2001pa}
                                                                                     \\ \hline
$F_{\pi\gamma\gamma^*}$, LCSR    &  $0.19 \pm 0.05$    & $0.12\pm 0.03$\,  \footnotesize{$(\mu=2.4)$} & \cite{Schmedding:1999ap} \\
$F_{\pi\gamma\gamma^*}$, LCSR    &  $0.32$  & $0.21$\,\footnotesize{$(\mu=2.4)$} & \cite{Bakulev:2002uc} \\
$F_{\pi\gamma\gamma^*}$, LCSR, R  & $0.44$     & $0.31$   & \cite{Bakulev:2005cp}\\
$F_{\pi\gamma\gamma^*}$, LCSR, R  & $0.27$     & $0.19$   & \cite{Agaev:2005rc}
                                                                                      \\ \hline
$F^{\rm em}_{\pi}$,LCSR        & $0.24\pm 0.14\pm 0.08$ &  $0.17\pm0.10\pm0.05$ & \cite{Braun:1999uj,Bijnens:2002mg}  \\
$F^{\rm em}_{\pi}$,LCSR, R      & $0.20\pm 0.03$         & $0.14\pm0.02$  & \cite{Agaev:2005gu}
                                                                                      \\ \hline
$F_{B\to\pi\ell\nu}$, LCSR       & $0.19\pm 0.19$         & $0.13\pm0.13$              & \cite{Ball:2005tb}
\\
$F_{B\to\pi\ell\nu}$, LCSR       & $0.16$                 & $0.11$        & \cite{Duplancic:2008ix}
                                                                                      \\ \hline
LQCD, {\footnotesize $N_f=2$}, CW      &   $0.289\pm 0.166$   & $0.201\pm 0.114$    &
{\footnotesize QCDSF/UKQCD}~\cite{Braun:2006dg}
 \\ LQCD, {\footnotesize $N_f=2\!+\!1$}, DWF   &    $0.334 \pm 0.129$  &  $0.233\pm 0.088$
& {\footnotesize RBS/UKQCD}~\cite{Donnellan:2007xr}
  \\ \hline
\end{tabular}
\end{center}
\caption[]{\sf The Gegenbauer moment $a_2(\mu^2)$.
 The CZ model involves $a_2 =2/3$ at the low scale $\mu=500$~MeV; for the discussion of the
 extrapolation to higher scales, see Ref.~\cite{Bakulev:2002uc}.  The abbreviations stand for:
QCDSR: QCD sum rules; NLC: non-local condensates; LCSR: light-cone sum rules; R: renormalon model
for twist-4 corrections; LQCD: lattice calculation;
$N_f=2(+1)$: calculation using  $N_f=2(+1)$ dynamical quarks; CW: non-perturbatively ${\mathcal O}(a)$
improved Clover--Wilson fermion action; DWF: domain
wall fermions. For convenience we present the results for two scales, $\mu=1$~GeV and
$\mu=2$~GeV, the relation is calculated in NLO.
 }
\label{tab:a2pi}
\renewcommand{\arraystretch}{1.0}
\end{table*}

Such calculations where pioneered in 1981 by Chernyak and Zhitnitsky  \cite{Chernyak:1981zz}
who derived the corresponding sum rule and obtained $a_2\sim 0.5$ at the scale of order
$\mu^2 = 1-1.5$~GeV$^2$. Extrapolating this number to a very low scale $\mu^2 =0.25$~GeV$^2$
and adding a simplifying assumption that higher-order coefficients $a_4,\ldots$ vanish, they have
formulated a simple model for the low-energy pion DA, which has become known as the
Chernyak-Zhitnitsky (CZ) model:
\begin{eqnarray}
  \phi^{\rm CZ}_\pi(x) &=& 30 x(1-x)(2x-1)^2
\nonumber\\ &=& 6 x(1-x)\left[1 +\frac23 C_2^{3/2}(2x-1)\right].
\label{eq:CZ}
\end{eqnarray}
This model corresponds to a very asymmetric momentum fraction distribution which vanishes
at the point where the pion momentum is shared equally between the quark and the
antiquark $\phi^{\rm CZ}_\pi(x=1/2)=0$. The striking difference of the CZ model and the
asymptotic DA has been fuelling an extensive and sometimes heated discussion for many years.

Newer estimates of $a_2$ following the CZ approach yield a somewhat smaller value
\cite{Khodjamirian:2004ga,Ball:2006wn}  $a_2(1$~GeV$)\sim 0.3$, the difference being due
to a combination of several factors: writing the sum rules for $a_2$ directly instead of
the second moment $\langle\xi^2\rangle$, taking into account the NLO corrections and
using slightly different values of the parameters.

Light-cone sum rules~\cite{Balitsky:1989ry,Braun:1988qv,Chernyak:1990ag} are
a modification of the general SVZ approach \cite{SVZ},
in which the pion DA serves as the main input in calculations of form factors (or hadron matrix elements).
The Gegenbauer coefficient $a_2$ is not calculated directly, but is
extracted from the comparison of the LCSR calculations with the experimental data. Note that in the case
of the pion transition form factor these fits are based on CLEO data~\cite{Gronberg:1997fj} only.
The results for $a_2$ are consistent with the direct calculations, see  Table~\ref{tab:a2pi}.

Finally, two independent lattice calculations of $a_2$ are now
available \cite{Braun:2006dg,Donnellan:2007xr}.
  The largest part of the uncertainty in these results is due to the chiral extrapolation.
 Overall, the results in  Table~\ref{tab:a2pi} show a consistent picture
\begin{eqnarray}
         a^{\rm NLO}_2(\mu^2=1~\mbox{GeV}^2) = 0.25\pm 0.10
\end{eqnarray}
and one can expect that the accuracy will increase in near future
when lattice calculations with physical pion mass become available.

Very little, unfortunately, is known about the next Gegenbauer coefficient, $a_4$.
The LCSR fits of heavy meson decay form factors indicate a small positive value,
$a_4\sim 0.04$ \cite{Duplancic:2008ix},
whereas the similar approach applied to pion transition form factor (CLEO data only) favors small
negative values \cite{Bakulev:2002uc,Bakulev:2005cp,Agaev:2005rc}.
Lattice calculations of this coefficient suffer from large (lattice) artifacts in the
operator renormalization and are not feasible at present.

The calculation of $a_4$ within the QCD sum rule approach has been
attempted in the so-called nonlocal condensate model (NLC)
\cite{Mikhailov:1991pt,Bakulev:1998pf,Bakulev:2001pa} which
involves a resummation of a tower of condensates of a certain
type. This approach leads to a sizeable negative value $a_4\sim
-0.1$ which is included in the so-called
Bakulev-Mikhailov-Stefanis (BMS) model of the pion DA. A large
negative value for $a_4$ in the NLC approach can be traced back to the
basic feature of this model that nonperturbative corrections to
the DA get smeared over a {\em finite} interval of momentum
fractions $\Delta x \sim \lambda_q^2 /(2M^2) \sim 0.2$ where
$\lambda^2_q = \langle \bar q D^2 q\rangle /\langle \bar q q
\rangle \sim 0.4$~GeV$^2$ is the average virtuality of quarks in
QCD vacuum and $M^2\sim 1$~GeV$^2$ is the Borel parameter. To our
opinion, appearance of $\Delta x$ is an artifact of the NLC model:
contributions of this type would produce ``bumps'' at large
values of Bjorken variable in quark parton distributions in the nucleon
\cite{Braun:1994jq} (which are absent) and also a finite smearing
proves not sufficient to cure the QCD sum rules for heavy-to-light
decay form factors \cite{Ali:1993vd} (which is the reason why this
technique was eventually abandoned and replaced by LCSRs). It
seems much more natural to assume that the nonperturbative
contributions get smeared over the whole interval of momentum
fractions $0<x<1$. A possible mechanism for such ``complete''
smearing is considered on the example of a photon DA in Appendix
B~in Ref.~\cite{Ball:2002ps}. We believe, therefore, that the
NLC-model-based predictions for $a_4$ have to be viewed with
caution.

Last but not least, we mention the LCSR calculation \cite{Braun:1988qv} for the pion DA
in the middle point:
\begin{equation}
   \phi_\pi(x=0.5, \mu^2 =1~\mbox{GeV}^2) = 1.2\pm 0.3\,,
\label{eq:middle}
\end{equation}
which is only available result beyond the Gegenbauer expansion.
This result excludes a large ``dip'' in the pion DA in the center
region and thus contradicts the CZ and BMS models. It is, however,
consistent with most of the parameterizations of the pion DA that
are used in vast literature on $B$-decays. Smaller values of
$\phi_\pi(x=0.5)$ are also strongly disfavored by numerous LCSR
calculations of pion-hadron couplings $g_{\pi NN},g_{\pi DD^*},
g_{\rho\omega\pi},\ldots$, see e.g.
\cite{Braun:1988qv,Belyaev:1994zk,Khodjamirian:1999hb,Aliev:1999ce,Aliev:2006xr,Aliev:2009kg}.

It has been suggested \cite{Radyushkin:2009zg,Polyakov:2009je} that the BaBar
data~\cite{BABAR} indicate an unusual ``flat'' DA $\phi_\pi(x)\simeq 1$ which does not
vanish at the end points so that the Gegenbauer expansion is not convergent
(more precisely: not uniformly convergent at the end points).
We believe that this conclusion is not warranted and in fact no conclusion on the end--point
behavior of the pion DA can be inferred on the basis of the existing experimental data.

\begin{figure}[t]
\centerline{
\begin{picture}(210,140)(0,0)
\put(-5,0){\epsfxsize7.8cm\epsffile{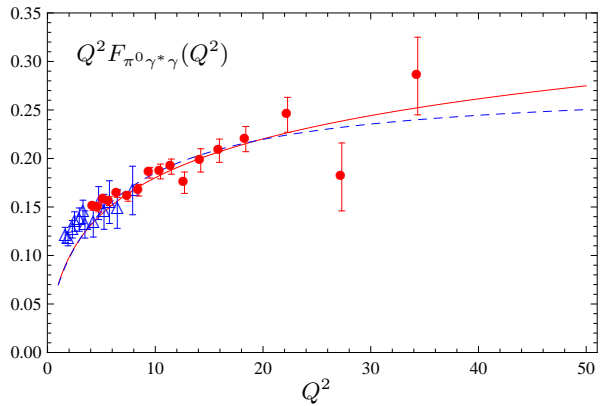}}
\put(105,-8){$Q^2$}
\put(20,120){$Q^2F_{\pi^0\gamma^*\gamma}(Q^2)$}
\end{picture}
}
\caption{Pion transition form factor in the MR model~(\ref{eq:MRmodel})
calculated using flat pion DA (solid red curve) and
CZ-type DA with $a_2=0.5$ and higher Gegenbauer moments set to zero
(dashed blue curve). The nonperturbative parameter
$\sigma = 0.53~\mbox{GeV}^2$ in both cases.
The experimental data are from~\cite{BABAR} (full circles)
and~\cite{Gronberg:1997fj} (open triangles).
}
\label{fig:MRmodel}
\end{figure}

To explain our statement, let us examine the argumentation in~\cite{Radyushkin:2009zg}
more closely. It is based on the elegant model for the transition form factor suggested
by Musatov and Radyushkin (MR) in an earlier work~\cite{Musatov:1997pu}, which
is derived from the exact two-body (e.g. quark-antiquark) contribution in the
noncovariant light-cone formalism of Brodsky and Lepage \cite{Lepage:1979zb}
under certain simplifying assumptions on the shape of the pion wave function:
\begin{equation}
    F^{\rm MR}_{\gamma^*\gamma \to \pi^0}(Q^2) = \frac{\sqrt{2}f_\pi}{3 Q^2}
   \int_0^1\! \frac{dx}{x} \phi_\pi(x)
  \left[1-\exp\left(-\frac{x Q^2}{2\bar x \sigma}\right)\right].
\label{eq:MRmodel}
\end{equation}
The first term in the square brackets corresponds to the usual leading-order contribution
to Fig.~\ref{fig:flow}a, whereas the second term is entirely a soft contribution
of the type Fig.~\ref{fig:flow}c. Note that this second term is exponentially small
in $Q^2$ for each finite quark momentum fraction, so it is absent in any order
of the operator product expansion. Using Eq.~(\ref{eq:MRmodel}) with
$\sigma = 0.53~\mbox{GeV}^2$~\cite{Radyushkin:2009zg}
and the ``flat'' pion DA $\phi_\pi(x)=1$ allows one to describe the apparent scaling violation
in the BaBar data, as illustrated by the solid curve in Fig.~\ref{fig:MRmodel}.

The caveat with this argument (and a very similar argumentation in Ref.~\cite{Polyakov:2009je})
is that flat DA is not necessary; in fact a CZ-type DA with $a_2=0.5$
$$
\phi_\pi(x) = 6 x(1-x)\left[1 +\frac12 C_2^{3/2}(2x-1)\right]
$$
yields a very similar (or even better) description of the data,
see the dashed (blue) curve in the same figure. Moreover, it is seen that the two DA models
can hardly be discriminated at all unless precise data with $Q^2 > 20$~GeV$^2$ are available!

It is easy to see why this happens. 
The flat DA $\phi^{\rm flat}_\pi(x) =1$ can be expanded
in Gegenbauer polynomials as follows:
\begin{equation}
 1 =\sum_{k=0,2,\ldots} a_k^{\rm flat}\varphi_k(x)\,,\qquad
a_k^{\rm flat} \,=\, \frac{2(2k+3)}{3(k+1)(k+2)}\,.
\label{eq:ak_flat}
\end{equation}
This equation has to be understood in the sense of distributions: The equality holds when both
sides are integrated with a test function that is finite (or does not increase too fast)
at the end points.

In perturbation theory (LO) the form factor is proportional to the sum of the Gegenbauer
coefficients
$$
 \int_0^1\! \frac{dx}{x} \phi_\pi(x)  = 3 \Big[1+a_2+a_4+\ldots\Big]\,.
$$
For the flat DA this series diverges, which motivates introduction of a certain
regulator, e.g. the soft correction given by second, exponential, term in Eq.~(\ref{eq:MRmodel})
or effective quark mass in Ref.~\cite{Polyakov:2009je}. Our observation is, however, that
{\it if} such a regulator is included, the Gegenbauer expansion for the form factor
is converging very rapidly and contributions of higher-order terms in this series
are in fact negligible. For illustration, consider the MR model
with a flat DA for, say $Q^2=20$~GeV$^2$, and check how much is being contributed by
each successive Gegenbauer polynomial. One obtains
\begin{eqnarray}
Q^2 F^{\rm MR,flat}_{\gamma^*\gamma \to \pi^0}(Q^2=20) &=& \frac{\sqrt{2}f_\pi}{3}\cdot 3.56513
\nonumber\\
&&\hspace*{-4cm}{}=
 \frac{\sqrt{2}f_\pi}{3}\Big[ 2.724 + 0.649 + 0.162 + 0.028 +\ldots \Big]
\end{eqnarray}
where the first term on the r.h.s. is the contribution of the asymptotic DA, the second
term is due to $a^{\rm flat}_2$ etc. One sees that all contributions beyond $n=4$ are tiny.

Our conclusion is that a good description of the BaBar data
\cite{BABAR} achieved in \cite{Radyushkin:2009zg,Polyakov:2009je}
is not due to an unusual end-point behavior of the DA, 
but rather to a (model dependent)
large nonperturbative soft correction to the form factor. Such a
large correction effectively suppresses contributions of higher
order terms in the Gegenbauer expansion and makes the question of
the end-point behavior of the pion DA irrelevant. The problem is,
therefore, whether such a large nonperturbative correction can be
expected, and whether it can be estimated in a model-independent
way. This is the question that we address in the next Section.

%
\section{Light Cone Sum Rules for the pion-photon transition}
%
\subsection{Dispersion approach}

The technique that we adopt in what follows has been suggested
originally by Khodjamirian in \cite{Khodjamirian:1997tk}.
It is to calculate the pion transition form factor for two large virtualities,
$Q^2$ and $q^2$, using the OPE, and make the analytic continuation to the real photon limit
$q^2=0$ using dispersion relations. In this way explicit evaluation of contributions
of the type in Fig.~\ref{fig:flow}b,c is avoided (since they do not contribute for sufficiently
large $q^2$) and effectively replaced by certain assumptions on the physical spectral density in the
$q^2$-channel.

The starting observation \cite{Khodjamirian:1997tk} is that $F_{\gamma^*\gamma^* \to \pi^0}(Q^2,q^2)$
satisfies an unsubtracted dispersion relation in the variable $q^2$ for fixed $Q^2$.
Separating the contribution of the lowest-lying vector mesons $\rho,\omega$ one can write
\begin{eqnarray}
 F_{\gamma^*\gamma^* \to \pi^0}(Q^2,q^2) &=& \frac{\sqrt{2}f_\rho F_{\gamma^*\rho \to \pi^0}(Q^2)}{m^2_\rho+q^2}
\nonumber\\&&\hspace*{-2cm}{} +
\frac{1}{\pi}\int_{s_0}^\infty ds\,\frac{\mathrm{Im} F_{\gamma^*\gamma^* \to \pi^0}(Q^2,-s)}{s+q^2}
\label{eq:DR}
\end{eqnarray}
where $s_0$ is a certain effective threshold. Here, the $\rho$ and $\omega$ contributions
are combined in one resonance term assuming $m_\rho\simeq m_\omega$ and the zero-width
approximation is adopted; $f_\rho\sim 200$~MeV is the usual vector meson decay constant.
Note that since there are no massless states, the real photon limit is recovered
by the simple substitution $q^2\to 0$ in (\ref{eq:DR}).

On the other hand, the same form factor can be calculated using QCD perturbation theory and the OPE.
The QCD result satisfies a similar dispersion relation
\begin{equation}
 F^{\rm QCD}_{\gamma^*\gamma^* \to \pi^0}(Q^2,q^2) = \frac{1}{\pi}\int_{0}^\infty ds\,\frac{\mathrm{Im} F^{\rm QCD}_{\gamma^*\gamma^* \to \pi^0}(Q^2,-s)}{s+q^2}\,.
\label{eq:DRQCD}
\end{equation}
The basic assumption of the method is that the physical spectral density
above the threshold $s_0$ coincides with the QCD spectral density as given by the OPE:
\begin{equation}
   \mathrm{Im}F_{\gamma^*\gamma^* \to \pi^0}(Q^2,-s) = \mathrm{Im}F^{QCD}_{\gamma^*\gamma^* \to \pi^0}(Q^2,-s)
\label{eq:duality}
\end{equation}
for $s>s_0$. This is the usual approximation of local quark-hadron duality.

We expect that the QCD result reproduces the ``true'' form factor for large values of $q^2$.
Equating the two representations (\ref{eq:DR}),(\ref{eq:DRQCD}) at $q^2\to-\infty$
and subtracting the contributions of $s>s_0$ from both sides one obtains
\begin{eqnarray}
  \sqrt{2}f_\rho F_{\gamma^*\rho \to \pi^0}(Q^2) = \frac{1}{\pi}\int_{0}^{s_0}\!\!ds\,
\mathrm{Im} F^{\rm QCD}_{\gamma^*\gamma^* \to \pi^0}(Q^2,-s)\,.
\end{eqnarray}
This relation explains why $s_0$ is usually referred to as the interval of duality (in the vector channel).
The (perturbative) QCD spectral density $\mathrm{Im}F^{QCD}_{\gamma^*\gamma^* \to \pi^0}(Q^2,-s)$ is a
smooth function and does not vanish at small $s\to 0$. It is very different from the physical
spectral density $\mathrm{Im}F_{\gamma^*\gamma^* \to \pi^0}(Q^2,-s) \sim \delta(s-m_\rho^2)$.
However, the integral of the QCD spectral density over a certain region of energies coincides
with the integral of the physical spectral density over the same region; in this sense the QCD
description of correlation functions in terms of quark and gluons is dual to the description in terms of
hadronic states.

In practical applications of this method one uses an additional trick, borrowed from
QCD sum rules \cite{SVZ}, which allows one to reduce the sensitivity on the duality
assumption in Eq.~(\ref{eq:duality}) and also suppress contributions of higher orders
in the OPE. The idea is essentially to make the matching between the ``true'' and
calculated form factor at a finite value $q^2 \sim 1-2$~GeV$^2$ instead of the $q^2\to\infty$
limit. This is done going over to the Borel representation $1/(s+q^2)\to \exp[-s/M^2]$
the net effect being the appearance of an additional weight factor under the integral
\begin{eqnarray}
  \sqrt{2}f_\rho F_{\gamma^*\rho \to \pi^0}(Q^2) &=& \frac{1}{\pi}\int_{0}^{s_0}ds\, e^{-(s-m^2_\rho)/M^2}\,
\nonumber \\&&{}\times
\mathrm{Im} F^{\rm QCD}_{\gamma^*\gamma^* \to \pi^0}(Q^2,-s)\,.
\label{eq:Frhogamma}
\end{eqnarray}
Varying the Borel parameter within a certain window, usually 1-2~GeV$^2$ one can test the
sensitivity of the results to the particular model of the spectral density.

With this refinement, substituting Eq.~(\ref{eq:Frhogamma}) in (\ref{eq:DR}) and using
Eq.~(\ref{eq:duality}) one obtains for $q^2\to 0$ \cite{Khodjamirian:1997tk}
\begin{eqnarray}
  F^{\rm LCSR}_{\gamma^*\gamma \to \pi^0}(Q^2) &\!=\!&
\frac{1}{\pi}\!\int_{0}^{s_0}\!\!\! \frac{ds}{m_\rho^2}
\mathrm{Im} F^{\rm QCD}_{\gamma^*\gamma^* \to \pi^0}(Q^2\!,-s)
e^{(m^2_\rho-s)/M^2} \nonumber\\&&{} +
\frac{1}{\pi}\int_{s_0}^\infty \frac{ds}{s} \mathrm{Im} F^{\rm
QCD}_{\gamma^*\gamma^* \to \pi^0}(Q^2,-s)\,. \label{eq:22}
\end{eqnarray}
This expression contains two nonperturbative parameters --- vector meson mass $m_\rho^2$ and effective threshold
$s_0\simeq 1.5$~GeV$^2$---
as compared to the ``pure'' QCD calculation, and the premium is that one does not need to evaluate
the contributions of Fig.~\ref{fig:flow}b,c explicitly: They are taken into account effectively via the
nonperturbative modification of the spectral density.

As an illustration, consider the leading twist QCD expression at
the leading order
\begin{equation}
   F^{\rm QCD}_{\gamma^*\gamma^*\to \pi^0}(Q^2,q^2) = \frac{\sqrt{2}f_\pi}{3}\,\int_0^1
 \frac{dx\,\phi_\pi(x)}{x Q^2+\bar x q^2}\,.
\label{eq:lolo}
\end{equation}
In this case the momentum fraction integral can easily be
converted to the form of a dispersion relation by the change of
variables $x\to s = Q^2\bar x/x$. The resulting LO and leading
twist LCSR expression is \cite{Khodjamirian:1997tk}
\begin{eqnarray}
  F^{\rm LCSR}_{\gamma^*\gamma\to \pi^0}(Q^2) &=&
 \frac{\sqrt{2}f_\pi}{3}\left\{
\int_0^{x_0}\!\! \frac{dx\,\phi_\pi(x)}{\bar x m_\rho^2}e^{(\bar x
m_\rho^2-xQ^2)/(\bar x M^2)} \right. \nonumber\\&&{} \left.
 +
 \frac{1}{Q^2}\int_{x_0}^{1} \frac{dx\,\phi_\pi(x)}{x} \right\}\,,
\label{eq:LCSR_LO}
\end{eqnarray}
where
\begin{equation}
 x_0 = \frac{s_0}{s_0+Q^2}\,.
\end{equation}
Note that the modification of the perturbative expression only
concerns the region $x<x_0\sim s_0/Q^2$ so this is a soft
contribution in our classification. Since $\phi_\pi(x) \sim x $
for $x\to 0$, this contribution (first term in (\ref{eq:LCSR_LO}))
corresponds to a power correction of the order of $s_0/Q^4$ for
$Q^2\to \infty$, in agreement with usual reasoning based on the
power counting.

\subsection{NLO perturbative corrections}

The NLO perturbative corrections to the $\gamma^*\gamma^*\pi^0$ form factor
for arbitrary photon virtualities were calculated in
Refs.~\cite{delAguila:1981nk,Braaten:1982yp,Kadantseva:1985kb}:
\begin{equation}
    F^{\rm QCD}_{\gamma^*\gamma^*\to \pi^0}(Q^2,q^2) =
\frac{\sqrt{2}f_\pi}{3}\!\!\!\int_0^1\!\!
 \frac{dx\,\phi_\pi(x)}{x Q^2\!+\!\bar x q^2}
\Big[1 + \frac{C_F\alpha_s}{2\pi} t(\bar x,w)\Big].
\end{equation}
The function $t(x,w)$ where  $w=Q^2/(Q^2+q^2)$
is given in Eq.~(5.2) in~\cite{Braaten:1982yp}.

Following Ref.~\cite{Schmedding:1999ap} we write the required imaginary part of
$F^{\rm QCD}_{\gamma^*\gamma^*\to \pi^0}(Q^2,q^2)$ as the sum of terms corresponding to
the expansion of the pion DA $\phi_{\pi}(x,\mu)$ in Gegenbauer polynomials (\ref{eq:C32}):
\begin{eqnarray}
 \lefteqn{\frac{1}{\pi}\mathrm{Im} F^{\rm QCD}_{\gamma^*\gamma^*\to \pi^0}(Q^2,-s) =}
\\
& = & \frac{\sqrt{2}f_\pi}{3}
\sum_{n=0}^\infty a_n(\mu)\left[\rho_n^{(0)}(Q^2,s) + \frac{C_F\alpha_s}{2\pi}\rho_n^{(1)}(Q^2,s;\mu)\right]
\nonumber
\end{eqnarray}
where the LO partial spectral density is proportional to the pion DA
\begin{equation}
\rho^{(0)}_{n}(Q^2,s)=\frac{\varphi_{n}(x)}{Q^2+s}\,, \qquad x = \frac{Q^2}{Q^2+s}\,.
\end{equation}
The NLO spectral density
\begin{eqnarray}
 \rho^{(1)}_{n}(Q^2,s) &=& \int_0^1dx\, \varphi_{n}(x)\frac{1}{\pi} \mathrm{Im}
\left[\frac{t(\bar x,w)}{x Q^2\!+\!\bar x q^2}\right]_{q^2=-s}
\end{eqnarray}
can be written in the following form:
\begin{widetext}
\begin{eqnarray}
\rho _{n}^{(1)}(Q^{2},s) &=&\frac{1}{2(Q^{2}+s)}\left\{
\left\{ -3\Big[1+2\big(\psi(2)-\psi(2+n)\big)\Big]
        +\frac{\pi ^{2}}{3}-\ln ^{2}\left( \frac{\bar x}{x}\right)
-\tilde\gamma^{(0)}_n
\ln \left( \frac{s}{\mu ^{2}}\right) \right\}
\varphi_{n}(x)\right.
\nonumber \\
&&\left.{}
+\tilde\gamma^{(0)}_n
\int_{0}^{\bar{x}}du\frac{\varphi _{n}(u)-\varphi _{n}(\bar{x})}{u-\bar{x}}
-2\left[ \int_{x}^{1}du\frac{\varphi_{n}(u)-\varphi _{n}(x)}{u-x}\ln \left( 1-\frac{x}{u}\right)
+(x\rightarrow \bar{x})\right] \right\},
\label{eq:ABOP}
\end{eqnarray}
\end{widetext}
where, as above, $x\equiv Q^2/(Q^2+s)$, $\psi(x)= d\ln \Gamma(x)/dx$
 and $\tilde\gamma^{(0)}_n$ is related to the
leading-order anomalous dimension $\gamma^{(0)}_n$ (\ref{eq:anomdim0}) as
\begin{equation}
 \gamma^{(0)}_n \equiv 2 C_F \tilde \gamma^{(0)}_n\,.
\end{equation}
Our result is similar but does not agree with the corresponding
expression in Ref.~\cite{Mikhailov:2009kf}. The difference is that
the first term in the second line in (\ref{eq:ABOP}) is not
symmetrized in $(x\rightarrow \bar{x})$ and hence the whole
expression in braces is not symmetric under this substitution. We
have checked that the spectral densities in (\ref{eq:ABOP})
reproduce the corresponding expressions for $n=0,2,4$
in~\cite{Schmedding:1999ap}: $\rho _{n}^{(1)}(Q^{2},s) = - A_n$.
We also checked by numerical integration for $n \leq 12$ that the
dispersion relation
\begin{eqnarray}
 \int_0^1 dx \frac{\varphi_n(x)\, t(\bar x,w)}{x Q^2\!+\!\bar x q^2}
= \int_0^\infty \!\frac{ds}{s+q^2}\, \rho _{n}^{(1)}(Q^{2},s)
\end{eqnarray}
is indeed satisfied.

As noticed in~\cite{Mikhailov:2009kf}, the integrals appearing in
Eq.~(\ref{eq:ABOP}) can be expanded in terms of $\varphi_n(x)$
with rational coefficients:
\begin{eqnarray}
&& \left[\int_{x}^{1}\!\!du\frac{\varphi_{n}(u)\!-\!\varphi _{n}(x)}{u-x}\ln \left(1\!-\!\frac{x}{u}\right)
+(x\rightarrow \bar{x})\right] =
\nonumber\\&&\hspace*{3.35cm}=-\sum_{k=0,2,\ldots}^n\!\! G_n^k\, \varphi_k(x)\,,
\label{eq:Gkn}
\end{eqnarray}
\begin{eqnarray}
&& \hspace*{-0.2cm}\int_{0}^{\bar{x}}\!\!du\frac{\varphi _{n}(u)-\varphi _{n}(\bar{x})}{u-\bar{x}}
 \,=\, -3\bar x+\!\!\! \sum_{k=0,1,\ldots}^n\!\! H_n^k\, \varphi_k(x)\,.
\label{eq:Hkn}
\end{eqnarray}
The matrices $G_n^k$ and $H_n^k$ can easily be calculated using orthogonality
relations for the Gegenbauer polynomials, e.g.
\begin{eqnarray}
 H_n^k &=& \mathcal{N}_k^{-1}\int_0^1 dx\,
 C_k^{3/2}(2x\!-\!1) \nonumber \\
 && \times \left [\int_{0}^{\bar{x}}du\frac{\varphi _{n}(u)-\varphi
 _{n}(\bar{x})}{u-\bar{x}}+3\bar{x} \right],
\end{eqnarray}
where $\mathcal{N}_k = (3/2)(k+1)(k+2)/(2k+3)$.
Explicit expressions for $n,k \le 12$ are collected in App.~B.

\subsection{Twist-four corrections}

%
\begin{figure*}[t]
\begin{center}
\includegraphics[width=.75\textwidth,
clip=true]{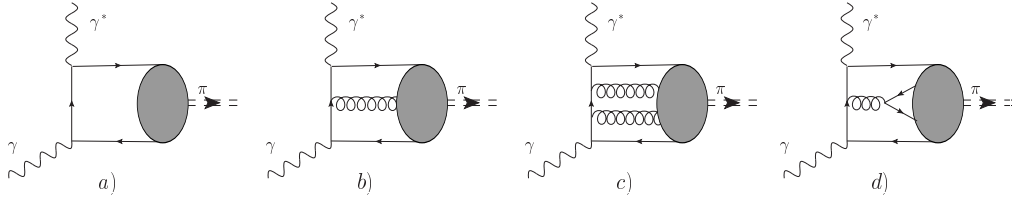}
\end{center}
\caption{Twist-4 corrections to the pion transition form factor}
\label{fig:twist4}
\end{figure*}

Twist-four corrections to the form factor, by definition, correspond to
the contributions of twist-four operators in the OPE of the time-ordered
product of the two electromagnetic currents in Eq.~(\ref{eq:Fgamma}).
Such contributions are of order $1/Q^4$, but, as we will see later, are not the
only ones that have to be taken into account to this accuracy: Twist counting
does not coincide in the present case with the counting of powers of large momentum,
so they should not be mixed.

Intuitively, twist-four effects can be thought of as due to quark
transverse momentum (or virtuality) in the handbag diagram shown
in Fig.~\ref{fig:twist4}a and quark-antiquark-gluon components in
the pion wave function, Fig.~\ref{fig:twist4}b. These two
contributions are related by exact QCD equations of motion
\cite{Braun:1989iv}; hence they must be taken into account
simultaneously. The corresponding calculation was done in
Ref.~\cite{Khodjamirian:1997tk}. The result (for two virtual
photons) can be written as
\begin{eqnarray}
 F_{\gamma^*\gamma^{*} \to \pi^0}(Q^2,q^2) &=& \frac{\sqrt{2}f_\pi}{3}
\Big(\int_0^1 dx \frac{ \phi_\pi(x)}{Q^2 x + q^2\bar x}
\nonumber\\
&&{} -  \int_0^1dx  \frac{\mathbb{F}_\pi(x)}{(Q^2 x + q^2\bar x)^2} \Big).
\label{eq:twist4}
\end{eqnarray}
and therefore
\begin{equation}
\frac{1}{\pi }\mathrm{Im} F_{\gamma^{\ast}\gamma^{\ast} \rightarrow
\pi ^{0}}(Q^{2},-s) =\frac{\sqrt{2}f_{\pi }}{3}\left[ \frac{\phi_{\pi }(x)}{s+Q^{2}}
-\frac{1}{Q^{2}}\frac{d\mathbb{F}_\pi(x)}{ds}\right]
\label{eq:3.1}
\end{equation}
with the usual substitution $x\equiv Q^{2}/(s+Q^{2})$.
The first term on the r.h.s. of (\ref{eq:twist4}), (\ref{eq:3.1})
is the leading order twist-two contribution
and ${\mathbb{F}_\pi(x)}$ is given in terms of the twist-4 pion DAs
\begin{widetext}
\begin{equation}
 \mathbb{F}_\pi(x,\mu) =  \frac{1}{4} \phi_{4;\pi}(x)
 + \int_0^{\bar x}\!\!\! d\alpha_1 \int_0^x\!\!\! d\alpha_2 \left[
 \frac{1}{\alpha_3}\left(\frac{x\!-\!\bar x \!+\!\alpha_1\!-\!\alpha_2}{\alpha_3}
 \Phi_{4;\pi}(\underline{\alpha}) - \widetilde \Phi_{4;\pi}(\underline{\alpha}) \right)
\right]_{\alpha_3=1-\alpha_1-\alpha_2}
\label{eq:mathF}
\end{equation}
\end{widetext}
where $\underline{\alpha} = \{\alpha_1,\alpha_2,\alpha_3\}$.
The two-particle twist-4 DA $\phi_{4;\pi}(x)$ is defined by the light-cone expansion
$y^2\to 0$ of the bilocal quark-antiquark operator~\cite{Braun:1989iv,Ball:1998je,Ball:2006wn}
\begin{eqnarray}
\lefteqn{\langle 0 | \bar q(0)\gamma_\mu\gamma_5 q(y)|\pi(p)\rangle =}
\nonumber\\
& = & i f_\pi p_\mu\! \int_0^1\!\! dx \, e^{-i x p\cdot y} \Big[\phi_{\pi}(x) +
\frac{y^2}{16} \phi_{4;\pi}(x)\Big] + \ldots
\label{eq:2pt}
\end{eqnarray}
and the three-particle DAs $\Phi_{4;\pi}(\underline{\alpha})$, $\widetilde\Phi_{4;\pi}(\underline{\alpha})$
correspond to the matrix elements
\begin{eqnarray}
\lefteqn{\langle 0 | \bar q(0)\gamma_\mu\gamma_5 gG_{\alpha\beta}(vn)q(un)|\pi(p)\rangle =}
\nonumber\\
& = & p_\mu (p_\alpha n_\beta -p_\beta n_\alpha)\, \frac{1}{pn}\, f_{\pi} \Phi_{4;\pi}(u,v;pn)
 + \dots,
\nonumber\\
\lefteqn{\langle 0 | \bar q(0)\gamma_\mu i g\widetilde{G}_{\alpha\beta}(vn)q(un)| \pi(p)\rangle=}
\nonumber\\
& = & p_\mu (p_\alpha n_\beta - p_\beta n_\alpha)\, \frac{1}{pn}\, f_\pi
\widetilde\Phi_{4;\pi}(u,v:pn)
 + \dots,
\label{eq:3pt}
\end{eqnarray}
with the shorthand notation
$${\cal F}(u,v;pn) = \int{\cal D}\underline{\alpha}\,e^{-ipn(u\alpha_1+v \alpha_3)} {\cal F}(\underline{\alpha}).$$
The Wilson lines in the definitions of the nonlocal operators in (\ref{eq:2pt}), (\ref{eq:3pt})
are not shown for brevity.
The integration measure is defined as
${\cal D}\underline{\alpha} = d\alpha_1d\alpha_2d\alpha_3 \delta(1-\alpha_1-\alpha_2-\alpha_3)$
and the dots denote contributions of other Lorentz structures that drop out
and also terms of twist 5 and higher. Our notation follows Ref.~\cite{Ball:2006wn}.

Strictly speaking, there exist also twist-4 contributions from the
wave function components containing two gluons or an extra
quark-antiquark pair, Fig.~\ref{fig:twist4}c and
Fig.~\ref{fig:twist4}d, but they are usually assumed to be small
and neglected. The relevant argument is based on the specific
property of four-particle twist-4 distributions: They do not allow
for a factorization in terms of two-particle distributions and,
say, quark or gluon condensate.

Higher-twist DAs can be studied using the conformal partial wave expansion, which is a
generalization of the Gegenbauer polynomial expansion for the leading twist DAs~\cite{Braun:1989iv}.
The contribution of the lowest conformal spin (asymptotic DAs) is
\begin{eqnarray}
  \mathbb{F}^{\rm as}_\pi(x,\mu) &=& \left(\frac{50}{3}+10\right) \delta_\pi^2(\mu) x^2(1-x)^2
\nonumber\\
 &=& \frac{80}{3} \delta_\pi^2(\mu) x^2(1-x)^2
\label{eq:t4_asympt}
\end{eqnarray}
where the first and the second contribution in the parenthesis are the contributions
of the two-particle and three-particle DAs in (\ref{eq:mathF}), respectively.
We stress that these two contributions are related by exact equations of motion;
taking into account e.g. the twist-4 correction to the handbag diagram and omitting
contributions of three-particle DAs is inconsistent with QCD. One can show \cite{Braun:2007wv}
that the next-to-leading order conformal spin contributions to the relevant DAs do not contribute
to $\mathbb{F}_\pi(x,\mu)$ so that this result is valid to NLO in the conformal expansion.
The coupling $\delta_\pi^2$ is defined by the matrix element
\begin{eqnarray}
&& \langle 0|\bar q g\widetilde G_{\mu\nu}\gamma^\nu q |\pi(p)\rangle
  = i p_\mu f_\pi \delta_\pi^2\,,
\nonumber\\
&& \delta_\pi^2(\mu) = \left(\frac{\alpha_s(\mu)}{\alpha_s(\mu_0)}\right)^{32/(9\beta_0)} \delta_\pi^2(\mu_0)\,.
\end{eqnarray}
This parameter was estimated using the QCD sum rule approach in
\cite{Novikov:1983jt} (see also \cite{Bakulev:2002uc}):
\begin{equation}
  \delta_\pi^2(\mu^2=1~\mbox{GeV}^2) \simeq 0.2~\mbox{GeV}^2\,.
\label{eq:delta2}
\end{equation}

\subsection{Twist-six corrections}

\begin{figure*}[t]
\begin{center}
\includegraphics[width=.65\textwidth,
clip=true]{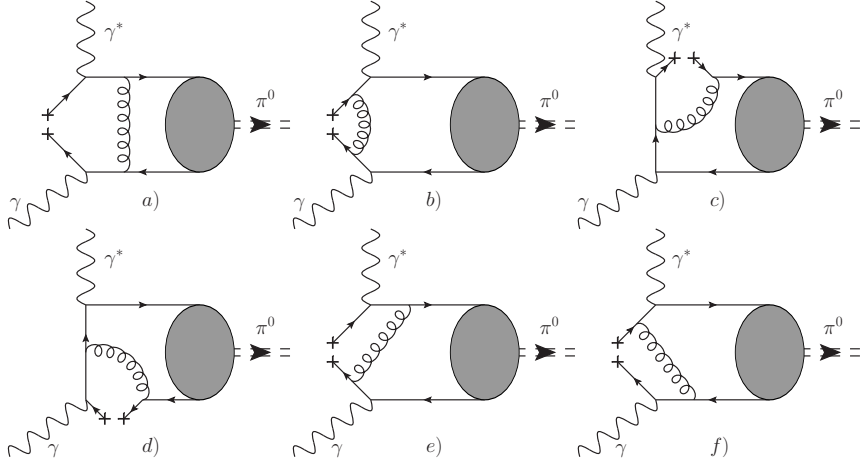}
\end{center}
\caption{Factorizable twist-6 corrections to the $F_{\gamma^*\gamma \to \pi^0}(Q^2)$
form factor.}
\label{fig:twist6}
\end{figure*}

The calculation of twist-six corrections to the transition form factor
presents a new result of this work.
To explain why such corrections may be important, consider an example
of the Feynman diagram shown in  Fig.~\ref{fig:twist6}a.
The broken quark line with crosses stands for
the quark condensate. If both photon virtualities are large, this diagram
contributes to the OPE of the product of the two electromagnetic currents
(the so-called cat-ears contribution) which involves the
twist-6 four-quark pion DA in the factorization approximation: One quark-antiquark
pair is put in the condensate and the other one forms a twist-3 chiral-odd quark-antiquark
DA. Explicit calculation gives (cf.~\cite{Khodjamirian:1997tk})
\begin{equation}
  F^{\rm Fig.\,4a}_{\gamma^*\gamma^{*} \to \pi^0}(Q^2,q^2) = \frac{\sqrt{2}f_\pi}{3}\,
\frac{32\pi\alpha_s\langle \bar q q\rangle^2 }{9f_\pi^2 q^2 Q^2}\int_0^1
 \frac{dx\,\phi^{p}_{3;\pi}(x)}{x Q^2+\bar x q^2},
\label{eq:6a}
\end{equation}
where~\cite{Braun:1989iv,Ball:1998je,Ball:2006wn}
\begin{equation}
\langle 0 | \bar q(0) i\gamma_5 q(\alpha n) | \pi(p)\rangle =
\frac{f_\pi m_\pi^2}{m_u+m_d}\, \int_0^1 dx \, e^{-ix \alpha pn}\,\phi^{p}_{3;\pi}(x)\,.
\label{eq:twist3}
\end{equation}
The DA $\phi^{p}_{3;\pi}(x)$ is related to the contribution of
three-particle (quark-antiquark-gluon) Fock state by equations of
motion~\cite{Braun:1989iv}. If the contributions of the twist-3
three-particle DA are neglected, $\phi^{p}_{3;\pi}(x)=1$ must be
taken~\cite{Braun:1989iv,Ball:1998je,Ball:2006wn}.
In
Eq.~(\ref{eq:6a}) we used the Gell-Mann-Oakes-Renner relation
$(m_u+m_d)(\langle \bar u u\rangle + \langle \bar d d\rangle) =
-f_\pi^2 m_\pi^2$ which is exact in the chiral limit.

For the case of two equal large virtualities $q^2=Q^2$ this
contribution is of order $\langle \bar q q\rangle^2/Q^6$, so it is
suppressed by two extra powers of $1/Q^2$ compared to the leading
term, as expected from dimension (twist) counting. On the other
hand, the real photon limit $q^2\to 0$ of (\ref{eq:6a}) cannot be
taken in a straightforward way since the pole at $q^2=0$ is
clearly unphysical. This singularity appears, obviously, because
the quark interacting with the soft photon comes close to the mass
shell. Thus the distance it travels becomes large and the OPE
cannot be applied. In the full theory, this singularity will be
tamed by nonperturbative corrections corresponding to photon
emission from large distances, Fig.~\ref{fig:flow}b. In the
simplest, vector meson dominance (VDM) approximation (in the LCSR
method in addition the continuum contribution is taken into
account) nonperturbative corrections amount to a replacement
$1/q^2\to 1/(m_\rho^2+q^2)$ so that for $q^2\to 0$ the singular
factor $1/q^2$ is replaced by $1/m_\rho^2$. One obtains
\begin{equation}
    F^{\rm Fig.\,4a}_{\gamma^*\gamma \to \pi^0}(Q^2)
 \simeq  \frac{\sqrt{2}f_\pi}{3}\,
\frac{32\pi\alpha_s\langle \bar q q\rangle^2 }{9 f_\pi^2 m_\rho^2 Q^4}
\int_0^1 \!dx\,\frac{\phi^{p}_{3;\pi}(x)}{x+\bar x m_\rho^2/Q^2}\,.
\label{eq:6a0}
\end{equation}
Note that this is a $1/Q^4$ correction to the form factor, not $1/Q^6$ as for equal
virtualities. The factor $1/m_\rho^2$ can be identified with the magnetic
susceptibility of the quark condensate (in the VDM approximation)
\cite{Ioffe:1983ju,Belyaev:1984ic,Balitsky:1985aq,Balitsky:1989ry,Ball:2002ps}
\begin{eqnarray}
  \chi \simeq \frac{2}{m_\rho^2} \simeq 3.3~\mbox{GeV}^{-2}.
\end{eqnarray}
which enters the definition of the leading twist DA of a real photon
\cite{Balitsky:1989ry,Ball:2002ps}
so that this contribution can be rewritten as a convolution of photon and pion DAs with the
coefficient function corresponding to a hard gluon exchange, cf.~Fig.~\ref{fig:flow}b.
The direct calculation of this contribution gives
\begin{eqnarray}
      F^{\rm Fig.\,1b}_{\gamma^*\gamma \to \pi^0}(Q^2) & = &  \frac{\sqrt{2}f_\pi}{3}\,
\frac{16\pi\alpha_s\chi \langle \bar q q\rangle^2 }{9 f_\pi^2 Q^4}
\nonumber\\&&{}\times \int_0^1 \!dx\,\frac{\phi^{p}_{3;\pi}(x)}{x}
\int_0^1 \!dy\,\frac{\phi_\gamma(y)}{\bar y^2}\,,
\label{eq:3b}
\end{eqnarray}
where $\phi_\gamma(y)\simeq 6 y(1-y)$ is the leading-twist photon DA \cite{Balitsky:1989ry,Ball:2002ps}.
The integrals over the quark momentum fractions in (\ref{eq:3b}) are both logarithmically
divergent at the end-points $x\to0$, $y\to1$, which signals that there is an overlap with the
soft region, Fig.~\ref{fig:flow}c.

Another twist-6 contribution comes from the diagram in Fig.~\ref{fig:twist6}b
\begin{equation}
F^{\rm Fig.\,4b}_{\gamma^*\gamma^*\to\pi^0}(Q^2,q^2)=\frac{\sqrt{2}f_\pi}{3}
\frac{64\pi\alpha_s\langle \bar q q\rangle^2 }{27f_\pi^2 }\!\int_0^1
 \!\!\frac{dx\,\phi^{\sigma}_{3;\pi}(x)}{(x Q^2\!+\!\bar x q^2)^3},
\end{equation}
where the DA $\phi^\sigma_{3;\pi}(x)$ is defined as
\begin{eqnarray}
\langle 0 | \bar q(0)\sigma_{\mu\nu} \gamma_5 q(\alpha n) | \pi(p)\rangle& =&
\frac{i}{6}(p_\mu x_\nu-p_\nu x_\mu)\frac{f_\pi m_\pi^2}{m_u+m_d}
\nonumber\\
&\times&\! \int_0^1\! dx \, e^{-ix \alpha pn}\,\phi^{\sigma}_{3;\pi}(x)\,.
\label{eq:twist3-2}
\end{eqnarray}
In the considered approximation (neglecting the quark-antiquark-gluon DAs)
the asymptotic form $\phi^\sigma_{3;\pi}(x)\,=\,6 x(1-x)$ must
be taken~\cite{Braun:1989iv,Ball:1998je,Ball:2006wn}.
Note that this contribution is also of the order of $1/Q^4$ and not $1/Q^6$ as suggested by
the naive power counting, since the limit $q^2\to 0$ leads to a quadratic divergence $\int dx/x^2$
at small momentum fractions. As above, this divergence is regulated in the LCSR approach
by correcting the spectral density to include the $\rho(\omega)$-resonance and the continuum.

Next, the diagrams in Figs.~\ref{fig:twist6}c,d can be calculated using the light-cone
expansion of the quark propagator in a background gluon field \cite{Balitsky:1987bk} and picking up
terms containing  covariant derivatives of the gluon field strength $D^\mu G_{\mu\nu}$.
These can be reduced to a quark-antiquark pair via equations of motion.
We have checked that there are no terms with additional derivatives compared to the expression given in
\cite{Balitsky:1987bk} contributing at the required twist 6 level.
A straightforward albeit rather lengthy calculation gives
\begin{widetext}
\begin{eqnarray}
F^{\rm Fig.\,4c,d}_{\gamma^*\gamma^*\to\pi^0}(Q^2,q^2)&=&-\dfrac{\sqrt{2}f_\pi}{3}\frac{16\pi\alpha_s\langle \bar q q\rangle^2 }{27f_\pi^2 }
\int_0^1dv(v-\bar{v})v \int_0^1du\dfrac{u-v}{\bar{u}}\phi^\sigma_{3;\pi}(u)
\nonumber\\
&\times&\left\{\dfrac{1}{[uvq^2+(1-uv)Q^2]^3}+\dfrac{1}{[(1-uv)q^2+uvQ^2]^3}\right\},
\end{eqnarray}
\end{widetext}
where it was used that to our accuracy
\begin{eqnarray}
\frac{x}{2}\left(\phi^p_{3;\pi}(x)+\frac{1}{6}\frac{d\phi^\sigma_{3;\pi}(x)}{dx}\right)&=&\frac{1}{6}\phi^\sigma_{3;\pi}(x)
\nonumber\\
\frac{\bar{x}}{2}\left(\phi^p_{3;\pi}(x)-\frac{1}{6}\frac{d\phi^\sigma_{3;\pi}(x)}{dx}\right)&=&\frac{1}{6}\phi^\sigma_{3;\pi}(x).
\end{eqnarray}
Finally, the diagrams in Figs.~\ref{fig:twist6}e,f vanish. This is
in difference to the similar calculation for the pion
electromagnetic form factor in Ref.~\cite{Braun:1999uj} where only
these diagrams contributed.

\subsection{The complete sum rule}

Collecting all contributions, we present here the complete light-cone
sum rule with twist-6 accuracy, which will be used in numerical
analysis in the next section.

The sum rule for the $\pi^0\gamma^*\gamma$ form factor can be written in terms
of the full QCD spectral density $\rho(Q^2\!,s)$ as
\begin{eqnarray}
F_{\gamma^\ast \gamma \to \pi ^0}(Q^2)
&=&\frac{\sqrt{2}f_{\pi}}{3}\Big[
\int_{s_{0}}^{\infty}\frac{ds}{s}\rho(Q^{2},s)
\nonumber\\
&&+
\frac{1}{m_{\rho }^{2}}\int_{0}^{s_{0}}\!\!ds\,\rho(Q^2\!,s)e^{(m_\rho^2-s)/M^2}
\Big]
\nonumber\\
&\equiv& F^{\rm hard}_{\gamma^\ast \gamma \to \pi ^0}(Q^2)
+ F^{\rm soft}_{\gamma^\ast \gamma \to \pi ^0}(Q^2)\,.
\label{eq:spectral24}
\end{eqnarray}
Here we define the ``hard'' and the ``soft'' contributions as
coming from large $s>s_0$ and small $s<s_0$ invariant masses in the
dispersion integral, respectively.  Note that the hard part is
model-independent whereas the soft part is obtained under the assumption
that the contribution of small invariant masses can be represented
by a single narrow resonance. The continuum threshold $s_0$ can be viewed
as the separation scale between hard and soft contributions; the dependence on
$s_0$ has to cancel in the sum.

The QCD spectral density, in turn, can be calculated as a sum of contributions
of different twists,  $t$=2,4,6
\begin{equation}
   \rho(Q^2\!,s) = \rho^{(2)}(Q^2\!,s) + \rho^{(4)}(Q^2\!,s) + \rho^{(6)}(Q^2\!,s) + \ldots
\end{equation}
The leading-twist spectral density to the NLO accuracy is given by
\begin{eqnarray}
\rho^{(2)}(Q^2,s)&=&
\frac{x}{Q^2}\sum_{n=0.2\ldots}^{\infty}a_{n}(\mu)\Bigg\{\varphi_{n}(x) +
\frac{C_F\alpha_{s}(\mu)}{4\pi}
\nonumber \\
&\times& \Bigg[ R_{n}(Q^{2},s)\varphi_{n}(x)
+ \tilde{\gamma}_{n}^{0}\sum_{k=0,1..}^{n}H_{n}^{k}\varphi_{k}(x)
\nonumber \\
&&+2\sum_{k=0,2..}^{n}G_{n}^{k}\varphi_{k}(x)
  -3 \tilde{\gamma}_{n}^{0}\bar x \Bigg] \Bigg\},
  \label{eq:55}
\end{eqnarray}
where
\begin{eqnarray}
R_n(Q^2,s)&=&-3\Big[1+2(\psi(2)-\psi(2+n)\Big]
\nonumber \\
&&{}+\frac{\pi^2}{3}-\ln^2\left(\frac{\bar {x}}{x}
\right)-\tilde{\gamma}_n^{(0)}\ln\left( \frac{s}{\mu^2}\right).
\end{eqnarray}
 The integrals
corresponding to the ``hard'' part, $s>s_0$, in
Eq.~(\ref{eq:spectral24}) can be taken analytically. The
corresponding expression in Eq.~(E.17) in~\cite{Bakulev:2002uc}
contains a misprint: the term $-3\ln^2(s_0/u)$ has to be replaced
by $-3\ln (s_0/u)$.

The twist-4 spectral density is equal to
\begin{equation}
\rho^{(4)}(Q^2,s)=\frac{160}{3Q^4}\delta_\pi^2(\mu)x^{3}\bar{x}(1-2x)\,.
\end{equation}
Finally, the twist-6 contribution can be written as:
\begin{eqnarray}
\rho^{(6)}(Q^{2}\!,s) &=& 8\pi C_F\alpha_s(\mu)\dfrac{\langle \bar{q} q\rangle^2}{N_c f_\pi^2}\dfrac{x^2}{Q^6}\Bigg[2x\log x+2x\log\bar{x}
\nonumber\\
&&{} - x + 2\delta(\bar{x})-\frac{1}{\bar{x}}+\delta(\bar{x})\int_{0}^1\frac{dx'}{\bar{x}'}\Bigg].
\label{eq:58}
\end{eqnarray}
Note that the last two terms combine to a ``plus'' distribution,
$1/[1-x]_+$. In all expressions (\ref{eq:55})--(\ref{eq:58})
$x\equiv Q^2/(Q^2+s)$.

We want to emphasize that the twist-6 contribution is not suppressed compared to the twist-4
one by an extra power of $Q^2$, and the same is true for all higher-twist corrections.
The twist expansion in LCSRs goes in powers of  $\Lambda^2_{\rm QCD}/s_0$,  $\Lambda^2_{\rm QCD}/M^2$
where $\Lambda^2_{\rm QCD}$ is a generic dimensionful parameter that characterizes
the size of higher-twist matrix elements.

%
\section{Numerical Analysis}
%

\subsection{The parameters}

All numerical results in this work are obtained using the two-loop running QCD
coupling with $\Lambda_{\rm QCD}^{(4)} = 326$~MeV and $n_f=4$ active flavors.
Unless stated otherwise, all nonperturbative parameters and models of the pion DA
refer to the renormalization scale $\mu_0=1$~GeV; $\alpha_s(\mu_0) = 0.494$.

A natural factorization and renormalization scale $\mu$ in the calculation
of the $\pi^0\gamma^*\gamma^*$ form factor with two large virtualities is given by the
virtuality of the quark propagator $\mu^2 \sim x Q^2 + \bar x q^2$ and depends on the quark momentum fraction.
If $q^2\to 0$, in the LCSR framework the relevant factorization
scale becomes $\mu^2 \sim x Q^2 + \bar x M^2$ or $\mu^2 \sim x Q^2 + \bar x s_0$
for large values of the Borel parameter, see e.g.~\cite{Braun:1999uj}.
Note that in the first integral in (\ref{eq:spectral24}) the quark virtuality is never
large, of order $Q^2$: The restriction $s < s_0$ translates to $\bar x < s_0/(s_0+Q^2)$
and hence $\mu^2 \simeq 2 s_0$ as $Q^2\to \infty$, in agreement with the interpretation
of this term as the ``soft'' contribution.
Numerical calculations with the $x$-dependent factorization
scale are rather slow so in this work we use a fixed scale, replacing $x$ by the constant
$\langle x \rangle$ which is varied within a certain range:
\begin{equation}
 \mu^2 = \langle x \rangle\, Q^2 + \langle \bar x \rangle\,s_0\,,
\qquad 1/4 < \langle x \rangle < 3/4\,.
\label{par:scale}
\end{equation}

The choice of the Borel parameter in LCSRs is discussed in~\cite{Ali:1993vd,Ball:1997rj}.
The subtlety is that the twist expansion in LCSRs goes in powers of $1/(x M^2)$
rather than $1/M^2$ in the classical SVZ approach. Hence one has to use somewhat larger
values of $M^2$ compared to the QCD sum rules for two-point correlation functions
in order to ensure the same hierarchy of contributions. We choose as the ``working window''
\begin{equation}
    1 < M^2 < 2~\text{GeV}^2
\end{equation}
and $M^2=1.5$~GeV$^2$ as the default value in our calculations.

We use  the standard value $s_0=1.5$~GeV$^2$ for the continuum threshold
as the central value, and the range
\begin{equation}
 1.3 < s_0 < 1.7~\text{GeV}^2
\end{equation}
in the error estimates.
We did not attempt to consider corrections due to the finite width
of the $\rho, \omega$ resonances. The estimates
in Ref.~\cite{Mikhailov:2009kf} suggest that such corrections
may result in the enhancement of the form factor by 2-4\% in
the small-to-mediate $Q^2$ region where the resonance part dominates.
We believe that such uncertainties are effectively covered by our
(conservative) choice of the continuum threshold.

Finally, we use the values $\delta_\pi^2=0.2\pm 0.4$~GeV$^2$ and
$\langle\bar q q \rangle = -(240\pm 10~\text{MeV})^3$
(at the scale 1 GeV) for the normalization parameter for
twist-4 DAs (\ref{eq:delta2}) and the quark condensate, respectively.

\subsection{Testing simple models}

We start our analysis with the comparison of the LCSR predictions
for the $\pi^0\gamma^*\gamma$ form factor for three simple models
of the pion DA that are often quoted in the literature:
\begin{eqnarray}
\phi_\pi^{\rm as}(x) &=& 6x(1-x)\,, \nonumber\\
 \phi_\pi^{\rm hol}(x) &=& \frac{8}{\pi}\sqrt{x(1-x)}\,,
\nonumber \\
\phi_\pi^{\rm flat}(x)&=& 1\,.
\label{eq:power}
\end{eqnarray}
The asymptotic $\phi_\pi^{\rm as}(x)$ and flat $\phi_\pi^{\rm
flat}(x)$ DAs have already been discussed above; the
``holographic'' model $\phi_\pi^{\rm hol}(x)$ is inspired by the
AdS/QCD correspondence \cite{Brodsky:2007hb} (see, however, \cite{Grigoryan:2008up}).

Apart from the general interest, considering these models allows one to test the
applicability of the Gegenbauer expansion. To this end, consider
the approximations to  $\phi_\pi^{\rm flat}(x)$, $\phi_\pi^{\rm hol}(x)$ by the truncated series
at order $n$:
\begin{eqnarray}
 \phi_\pi^{\rm flat(\rm hol),(n)}(x) &=&\sum_{k=0,2,\ldots}^n a_k^{\rm flat (\rm hol)}
 \varphi_{k}(x)\,,
\label{eq:trunc}
\end{eqnarray}
where
\begin{equation}
 a_k^{\rm hol} = \frac{2n+3}{3\pi}\left(\frac{\Gamma[(n+1)/2]}{\Gamma[(n+4)/2]}\right)^2
\end{equation}
and $a_k^{\rm flat}$ are given in Eq.~(\ref{eq:ak_flat}).

\begin{figure}[t]
\begin{center}
\begin{picture}(210,140)(0,0)
\put(0,0){\epsfxsize7.3cm\epsffile{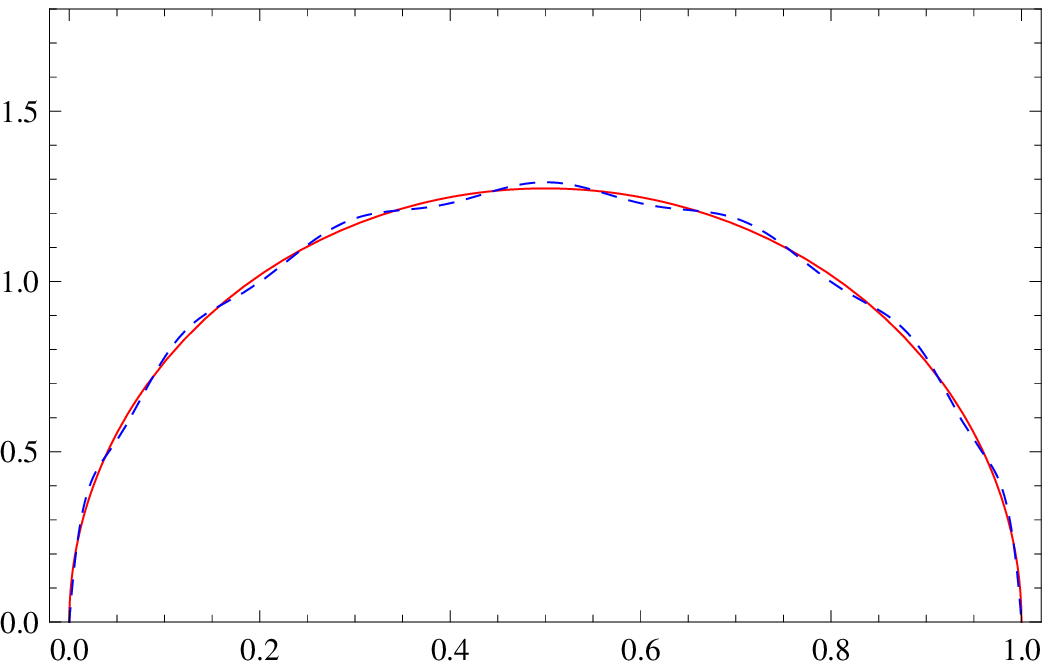}}
\put(105,-8){$x$}
\put(20,110){$\phi_\pi^{\rm hol}(x)$}
\end{picture}
\\[2mm]
\begin{picture}(210,140)(0,0)
\put(0,0){\epsfxsize7.3cm\epsffile{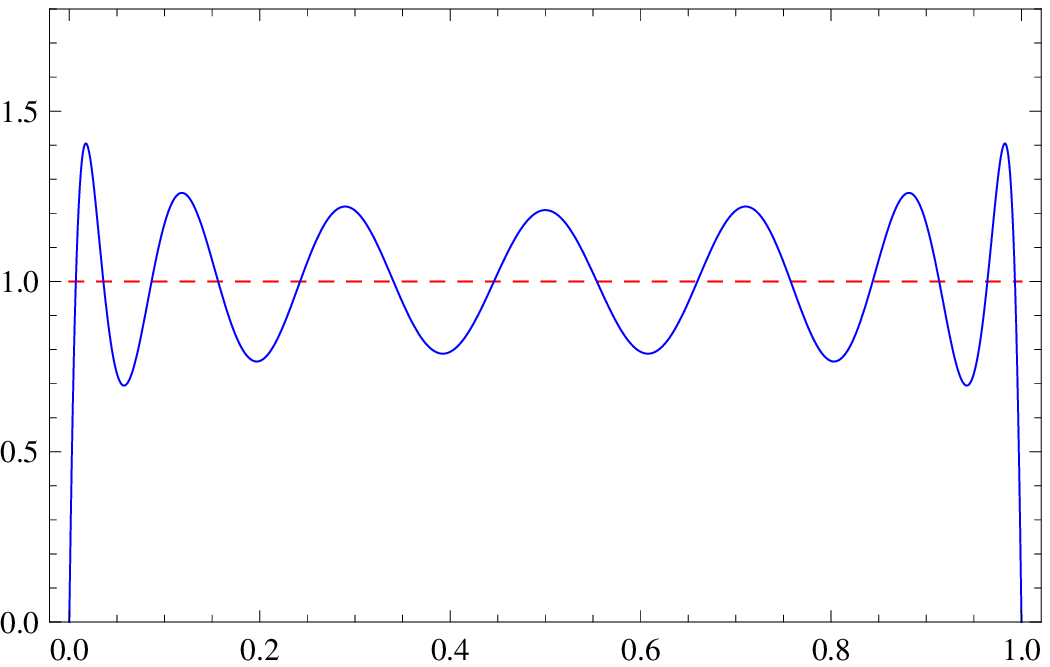}}
\put(105,-8){$x$}
\put(20,110){$\phi_\pi^{\rm flat}(x)$}
\end{picture}
\end{center}
\caption{The $n=12$ truncations (\ref{eq:trunc}) of the pion DAs
$\phi_\pi^{\rm hol}(x)$ and $\phi_\pi^{\rm flat}(x)$ compared to the exact
expressions.
}
\label{fig:DAflat+hol}
\end{figure}

The expressions collected in Sect.~3 and App.~A,B allow us to construct the sum rules
using up to seven terms $k=0,2,\ldots,12$ corresponding to the $n=12$ truncation.
The resulting DAs $\phi_\pi^{\rm flat,(n=12)}(x)$, $\phi_\pi^{\rm hol,(n=12)}(x)$ are compared with the exact ones,
$\phi_\pi^{\rm flat}(x)$, $\phi_\pi^{\rm hol}(x)$, in Fig.~\ref{fig:DAflat+hol}. Note that the $n=12$
approximation is very good for the ``holographic'' DA, whereas for the ``flat'' one the convergence
is slow and there are large oscillations.

\begin{figure}[t]
\centerline{
\begin{picture}(210,140)(0,0)
\put(-5,0){\epsfxsize7.8cm\epsffile{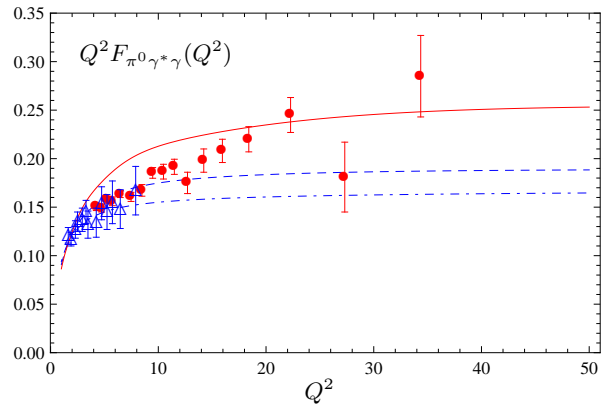}}
\put(105,-8){$Q^2$}
\put(20,120){$Q^2F_{\pi^0\gamma^*\gamma}(Q^2)$}
\end{picture}
}
\caption{
The pion transition form factor for the ``flat'' (solid red line),
``holographic'' (dashed blue line) and ``asymptotic'' (dash-dotted blue line) models
for the pion DA, cf. Eq.~(\ref{eq:power}).
The experimental data are from~\cite{BABAR} (full circles)
and~\cite{Gronberg:1997fj} (open triangles).
}
\label{fig:3DAs}
\end{figure}

\begin{figure}[t]
\centerline{
\begin{picture}(210,140)(0,0)
\put(-5,0){\epsfxsize7.8cm\epsffile{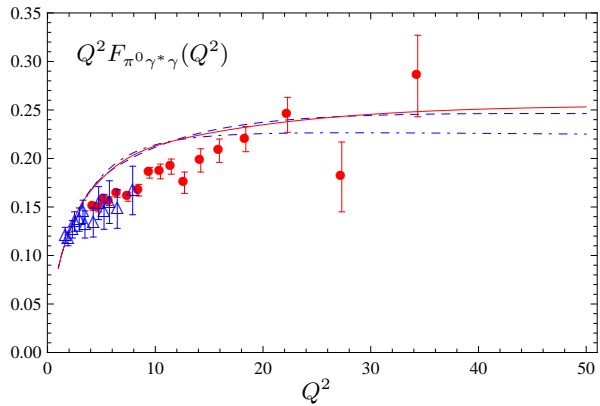}}
\put(105,-8){$Q^2$}
\put(20,120){$Q^2F_{\pi^0\gamma^*\gamma}(Q^2)$}
\end{picture}
} \caption{ The pion transition form factor for three different
approximations for the ``flat'' DA: $n=12$ (solid red line), $n=8$
(dashed blue line), and $n=4$ (dash-dotted). The experimental data
are from~\cite{BABAR} (full circles) and~\cite{Gronberg:1997fj}
(open triangles). } \label{fig:ConstDA}
\end{figure}

Next, we use the three DAs in Eq.~(\ref{eq:power}) (the last two ones truncated at order $n=12$)
to calculate the $\pi^0\gamma^*\gamma$ form factor.
The results are shown in Fig.~\ref{fig:3DAs}.
One sees that both the asymptotic and holographic models fail to describe the
BaBar data~\cite{BABAR}. The flat DA fares better for the largest $Q^2$ values, but
is considerably above the experiment at intermediate $Q^2\sim5-15$~GeV$^2$. In order to understand
this behavior, we compare in Fig.~\ref{fig:ConstDA} the predictions for three different
truncations of the flat DA: $\phi_\pi^{\rm flat,(n=12)}(x)$,  $\phi_\pi^{\rm flat,(n=8)}(x)$ and
$\phi_\pi^{\rm flat,(n=4)}(x)$.
We observe that all three calculations are very close to each
other for $Q^2 \leq 18$~GeV$^2$ so that in this region
contributions of Gegenbauer polynomials starting with $n=6-8$ play
no role. This conclusion is in agreement with our discussion of
the MR model in Sect.~2 and also supports the usual procedure of
modelling the pion DA by the asymptotic expression and
contributions of first two Gegenbauer polynomials in applications
to $B$-decays and pion electromagnetic form factor,
e.g.~\cite{Khodjamirian:1997tk,Schmedding:1999ap,Bakulev:2001pa,%
Bakulev:2002uc,Bakulev:2003cs,Agaev:2005rc,Mikhailov:2009kf}. We
also see that contributions of the Gegenbauer polynomials
$n=6\,,8$ become significant for the momentum transfers $Q^2>
18\,{\rm GeV}^2$. In the region $Q^2> 30\,{\rm GeV}^2$
higher-order polynomials should be included into analysis as well,
but the accuracy of the existing experimental data is not
sufficient to draw definite conclusions.

In other words, the differences between the predictions of
asymptotic, holographic and flat DAs in Fig.~\ref{fig:3DAs} for
$Q^2 < 20~{\rm GeV}^2$ are mostly due to the different
values of the coefficients $a_2$ and $a_4$, with 
$a_6$ also playing some role. This leaves us with $2-3$ parameters
that can be tuned to attempt a better description of the BaBar
data, the task that we address now.

\subsection{Confronting the BaBar data}

The extraction of the pion DA with
meaningful error estimates requires a global fit to the pion transition
and electromagnetic form factor, weak $B(D)\to \pi \ell \nu$ decays and
the couplings $g_{\pi NN}, g_{BB^*\pi}$ etc. using a Monte Carlo
scan of the space of all available parameters, 
which goes beyond the tasks of this work.
Fitting of the BaBar data is not attempted. Instead, we
present results for three sample models that describe the
$\pi^0\gamma^*\gamma$ form factor sufficiently well and discuss
their general features.

\begin{figure}[t]
\begin{center}
\begin{picture}(210,140)(0,0)
\put(0,0){\epsfxsize7.3cm\epsffile{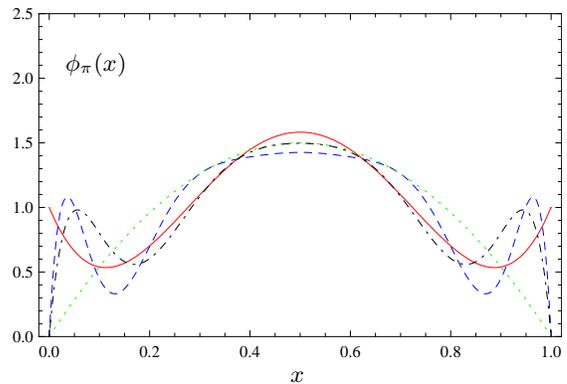}} \put(105,-8){$x$}
\put(20,110){$\phi_\pi(x)$}
\end{picture}
\end{center}
\caption{Model I, 
(red solid curve), model
II (blue dashed curve) and model III (black dash-dotted curve) of
the pion DA at the scale 1 GeV. The asymptotic pion DA is shown by
green dots for comparison. } \label{fig:DAmodels}
\end{figure}

The three models that we consider below are shown in
Fig.~\ref{fig:DAmodels} (at the scale 1 GeV) and the corresponding
Gegenbauer coefficients are collected in Table~\ref{tab:models}.

The first model
\begin{equation}
\phi_{\pi}^{\rm I}(x)=1-(7/18-0.13)\varphi_2(x) \label{eq:modelI}
\end{equation}
is nothing but the flat DA with the reduced second Gegenbauer
coefficient, $a^{\rm flat}_2 \to 0.130$.
\begin{table}[tb]
\renewcommand{\arraystretch}{1.3}
\begin{center}
\begin{tabular}{|c|c|c|c|c|c|c|c|} \hline
  Model & scale & $a_2$ & $a_4$ & $a_6$ & $a_8$ & $a_{10}$ & $a_{12}$
\\ \hline
 \multirow{2}{*}{  I } & $\mu=1$~GeV
& 0.130 & 0.244 & 0.179 & 0.141 & 0.116 & 0.099
\\
                            & $\mu=2$~GeV
& 0.089 & 0.148 & 0.097 & 0.070 & 0.054 & 0.044
\\ \hline
 \multirow{2}{*}{ II } & $\mu=1$~GeV
& 0.140 & 0.230 & 0.180 & 0.05 & 0.0 & 0.0
\\
                            & $\mu=2$~GeV
& 0.096 & 0.140 & 0.098 & 0.024 & -0.001 & $ < 10^{-3}$
\\ \hline
 \multirow{2}{*}{ III } & $\mu=1$~GeV
 & 0.160 & 0.220 & 0.080 & 0.0 & 0.0 & 0.0
\\
                            & $\mu=2$~GeV
 & 0.110 & 0.133 & 0.043 & -0.001 & $ < 10^{-3}$ & $ < 10^{-3}$
\\ \hline
\end{tabular}
\end{center}
\caption[]{\sf Gegenbauer coefficients of three sample models of pion DA
that are consistent with BaBar measurements \cite{BABAR} of the transition form factor,
cf. Fig.~\ref{fig:BABARfit}.} \label{tab:models}
\renewcommand{\arraystretch}{1.0}
\end{table}
It has a long ``tail'' of higher-order Gegenbauer polynomials.
From the previous discussion we expect that this ``tail'' actually
gives no contribution in the $Q^2$ range of interest. In order to
check that this is indeed the case, we consider the second model
in which the higher-order coefficients $a_{10}, a_{12}$ are put to
zero at the reference scale 1 GeV, and we keep a small $a_{8}$
to avoid an oscillating behavior at $x \sim 1/2$. 
Note that nonzero values of the coefficients
$a_{10}, a_{12}$ (and all higher) are generated at higher scales,
but this mixing is numerically insignificant. Finally, the third
model is chosen to explore the sensitivity of the predictions to
particular values of $a_2,\,a_4$ and $a_6$, and to see whether
they are correlated.

\begin{figure}[t]
\begin{center}
\begin{picture}(210,140)(0,0)
\put(-5,0){\epsfxsize7.8cm\epsffile{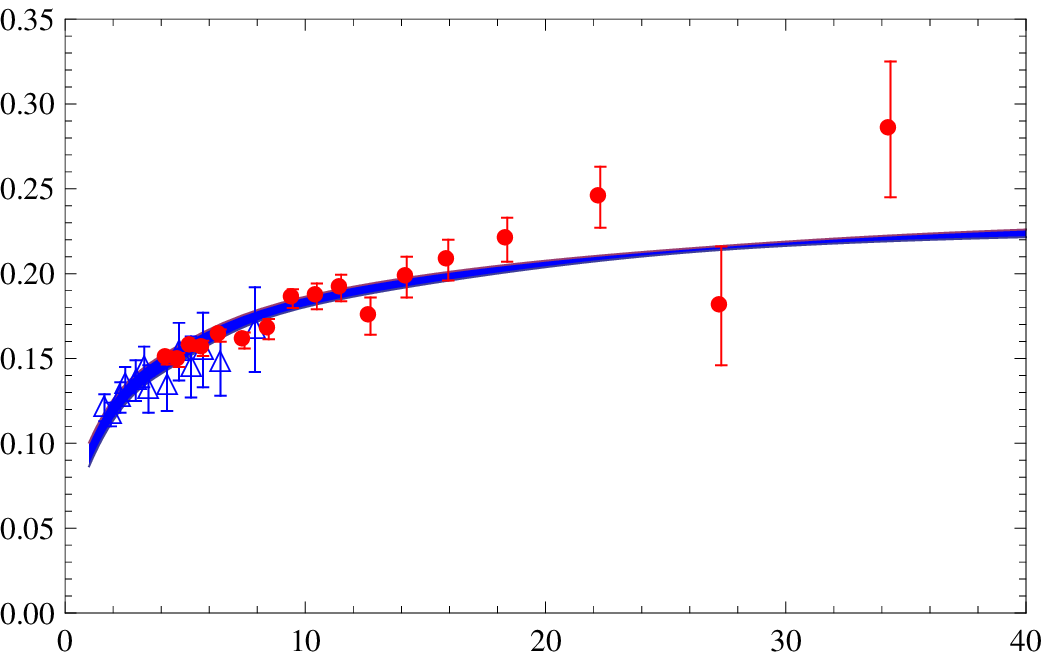}}
\put(105,-8){$Q^2$}
\put(20,120){$Q^2F_{\pi^0\gamma^*\gamma}(Q^2)$}
\put(170,15){Model I}
\end{picture}\\[3mm]
\begin{picture}(210,140)(0,0)
\put(-5,0){\epsfxsize7.8cm\epsffile{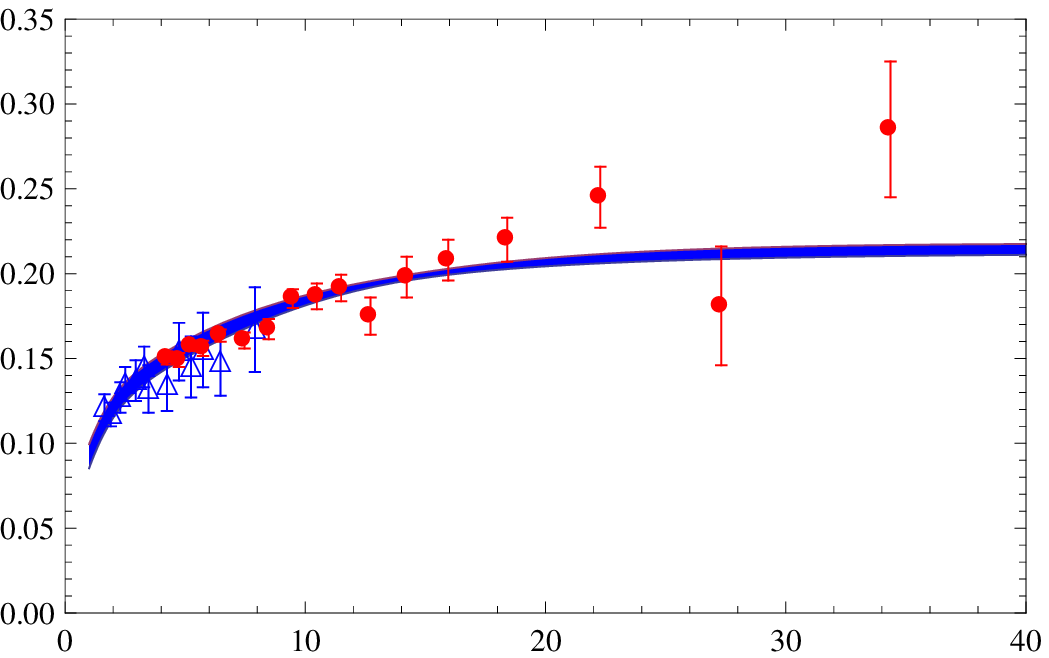}}
\put(105,-8){$Q^2$}
\put(20,120){$Q^2F_{\pi^0\gamma^*\gamma}(Q^2)$}
\put(170,15){Model II}
\end{picture}\\[3mm]
\begin{picture}(210,140)(0,0)
\put(-5,0){\epsfxsize7.8cm\epsffile{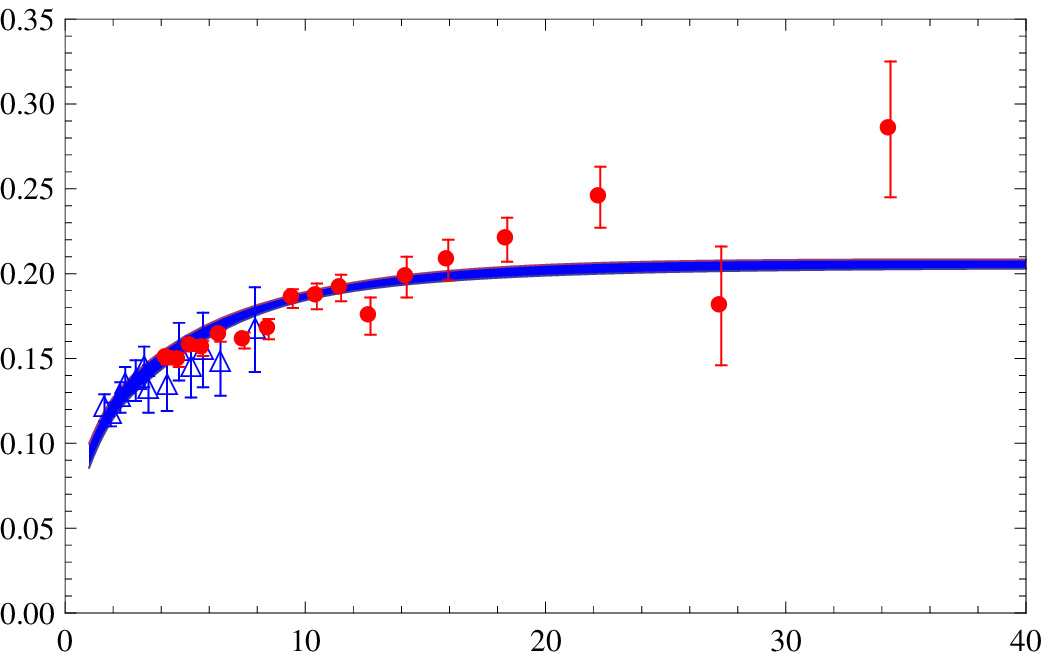}}
\put(105,-8){$Q^2$}
\put(20,120){$Q^2F_{\pi^0\gamma^*\gamma}(Q^2)$}
\put(170,15){Model III}
\end{picture}
\end{center}
\caption[]{\sf
The pion transition form factor for the three models of the pion
DA specified in the text.
The experimental data are from~\cite{BABAR} (full circles)
and~\cite{Gronberg:1997fj} (open triangles).
}
\label{fig:BABARfit}
\end{figure}

The calculations using these models are compared with the available experimental data
in Fig.~\ref{fig:BABARfit}. The results are shown by thick solid curves:
The line thickness shows the uncertainty and is calculated as a square root of the
sum of squares of the error bars on the LCSR predictions due to
variation of the parameters within the limits specified in Sect.~IV.A.
These include the factorization scale dependence,
dependence on the Borel parameter $M^2$ and the continuum threshold $s_0$,
and on higher-twist parameters $\delta_\pi^2$ and $\langle \bar q q \rangle$.

The distinctive feature of all three models is the large value
of the fourth Gegenbauer moment, $a_4$, which is necessary
in order to accommodate the observed rise of the scaled form factor in the
$Q^2 = 5 - 20$~GeV$^2$ range. It is not possible to trade the large
value of $a_4$ for the increased $a_2$ or $a_6$, although of course there
is some correlation.

Our value of $a_2\sim 0.13-0.16$ at 1~GeV is at the low end of the 
existing estimates, cf. Table.~\ref{tab:a2pi}, and in particular it is
lower compared to the earlier LCSR analysis of the 
transition form factor in
Refs.~\cite{Bakulev:2002uc,Mikhailov:2009kf,Agaev:2005rc}. The main
reason for this difference is that the Borel parameter in
\cite{Bakulev:2002uc,Mikhailov:2009kf} is fixed at
an {\it ad hoc}\ value $M^2=0.7\, {\rm GeV}^2$, whereas in the
present analysis we allow its variation in
the $1-2$~GeV$^2$ range. For the specific choice
$a_2(\mu_{\rm SY})=0.14$, $a_4(\mu_{\rm SY})=-0.09$, $\mu_{\rm
SY}\simeq 2.4$~GeV, advocated in
\cite{Bakulev:2002uc,Mikhailov:2009kf}, the result for the form
factor at $Q^2=5$~GeV$^2$ is increased by $\sim 11\%$ if $M^2$ is
changed from 0.7 to 1.5~GeV$^2$. Another reason is that in
\cite{Bakulev:2002uc,Mikhailov:2009kf,Agaev:2005rc} the twist-6
correction is not included. The size of this correction depends
strongly on the Borel parameter. For our choice $M^2\sim 1.5\pm
0.5$~GeV$^2$ the twist-6 term proves to be small: factor three
smaller that the twist-4 correction (see below), which is
gratifying as it signals convergence of the OPE. In contrast, at
$M^2=0.7$~GeV$^2$ the twist-6 correction is almost of the same
size as twist 4 and has opposite sign. Hence it must be included.
In both cases (increasing the Borel parameter and/or including the
twist-6 correction) the net effect is the increase of the form
factor by 5-10\% in the CLEO range which has to be compensated by
a smaller value of the second Gegenbauer moment.

The error band indicated by thickness of the curves in
Fig.~\ref{fig:BABARfit} has to be taken with caution. A  weak
scale dependence of our results is largely due to strong
cancellations of the NLO radiative corrections between the
contributions of the asymptotic DA and higher Gegenbauer
polynomials and may not be representative for the size of NNLO
corrections which are only known in the $\overline{\mbox{\rm CS}}$
factorization scheme, see \cite{Melic:2002ij} for a detailed
discussion of the related ambiguities. Also the uncertainty in the
twist-4 contribution is not reduced to the $\delta_\pi^2$
parameter: Using an alternative, renormalon model
\cite{Braun:2004bu} of the twist-4 pion DA generally produces
somewhat larger corrections. We have checked that the difference
is not very significant, however, and does not affect any of our
conclusions. Hence we do not show the corresponding results.

\begin{figure}[t]
\centerline{
\begin{picture}(210,140)(0,0)
\put(-5,0){\epsfxsize7.8cm\epsffile{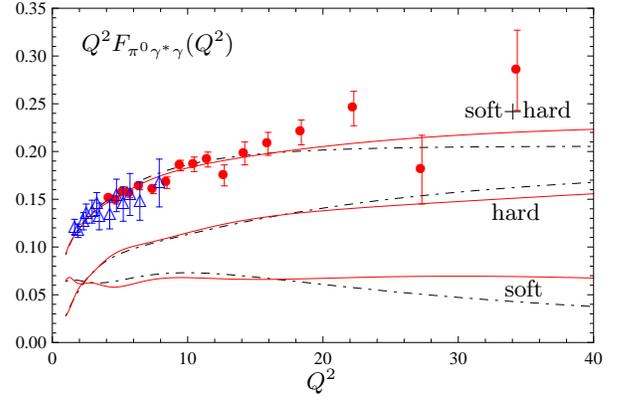}}
\put(105,-8){$Q^2$}
\put(20,120){$Q^2F_{\pi^0\gamma^*\gamma}(Q^2)$}
\put(180,27){soft}
\put(175,56){hard}
\put(165,95){soft+hard}
\end{picture}
}
\caption{
Contributions to the  $\pi^0\gamma^*\gamma$ form factor
from large (``hard'') and small (``soft'') invariant masses
in the dispersion representation, cf. Eq.~(\ref{eq:spectral24}),
for model I (solid curves) and model III (dash-dotted curves).
The experimental data are from~\cite{BABAR} (full circles)
and~\cite{Gronberg:1997fj} (open triangles).
}
\label{fig:softhard}
\end{figure}

The ``hard'' and ``soft'' contributions to the $\pi^0\gamma^*\gamma$
form factor as defined in Eq.~(\ref{eq:spectral24}) are shown separately for model I
(solid curves) and model III (dash-dotted curves) in Fig.~\ref{fig:softhard}.
Asymptotically, for $Q^2\to\infty$, the soft contribution is
power-suppressed compared to the hard one, $\sim s_0/Q^2$. This
suppression sets in for very large values of $Q^2$, however,
especially if the pion DA is enhanced close to the end points.
E.g. for our model III the soft contribution still accounts for
ca. 25\% of the form factor at $Q^2=30$~GeV$^2$ (for the
separation scale $s_0=1.5$~GeV$^2$). This means that a purely
perturbative leading twist QCD calculation of the transition form
factor for one real photon in collinear factorization should not
be expected to have high accuracy. A lattice calculation of the
transition $\pi\rho\gamma^*$ form factor at $Q^2\sim 2-5$~GeV$^2$
would help to estimate the contribution of the resonance region
more reliably.

\begin{figure}[t]
\centerline{
\begin{picture}(210,140)(0,0)
\put(-5,0){\epsfxsize7.8cm\epsffile{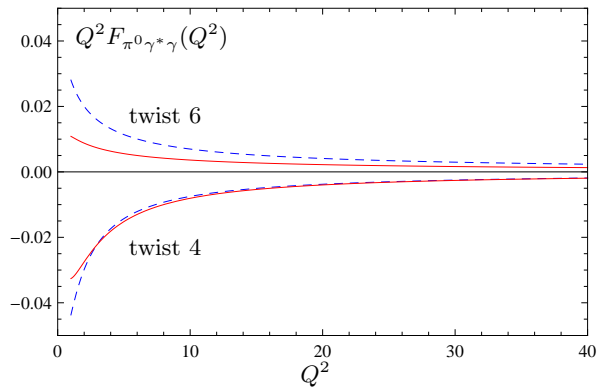}}
\put(105,-8){$Q^2$}
\put(20,120){$Q^2F_{\pi^0\gamma^*\gamma}(Q^2)$}
\put(40,40){twist 4}
\put(40,90){twist 6}
\end{picture}
} \caption{Higher twist contributions to the $\pi^0\gamma^*\gamma$
form factor for the values of the Borel parameter $M^2=1.5$ (solid
curves) and $M^2=0.7$~GeV$^2$ (dashed curves). }
\label{fig:twist46}
\end{figure}

Finally, in Fig.~\ref{fig:twist46} we show the higher-twist
contributions. The twist-4 correction is negative and the twist-6
one is positive. It turns out that the twist-6 contribution
depends rather strongly on the Borel parameter. It is suppressed
in the $Q^2$ region of interest relative to the twist-4 term for
our choice $M^2=1.5\,{\rm{GeV}}^2$, but increases rapidly for
smaller $M^2$. For example, for $M^2=0.7\, {\rm{GeV}}^2$ used in
\cite{Bakulev:2002uc,Mikhailov:2009kf}, the twist-6 correction is
$\sim 0.6$ of the twist-4 term at $Q^2=1\,{\rm{GeV}}^2$, becomes
equal (with opposite sign) at $Q^2 \simeq 14 \,{\rm{GeV}}^2$ and
overshoots twist-4 for larger $Q^2$ (because it contains a
logarithmic $\sim \ln Q^2$ enhancement).
%

\subsection{Other processes}
The pion DA is a universal function and, if extracted from one 
reaction, should, in general, describe all exclusive or
semi-inclusive processes that involve a pion in the initial and/or
final states. The most prominent of them are the pion electromagnetic 
form factor and the weak semileptonic decay rate  
$B \to \pi\ell\nu_\ell$. Without going in detail, we present here the 
corresponding LCSR calculations using the pion DA models as 
specified above.

The LCSRs for the pion electromagnetic form factor were derived in
Refs.~\cite{Halperin,Braun:1999uj,Bijnens:2002mg} and later
explored also in \cite{Agaev:2005gu,Agaev:2008}. 
These sum rules are known to the same accuracy
as for the transition form factor, i.e. including the NLO 
perturbative contribution, twist-4 and twist-6 corrections. 
Explicit expressions can be found in \cite{Bijnens:2002mg}.

\begin{figure}[t]
\centerline{
\begin{picture}(210,140)(0,0)
\put(-5,0){\epsfxsize7.8cm\epsffile{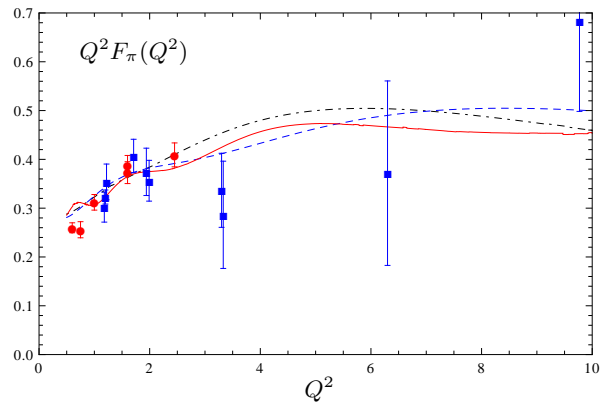}}
\put(105,-8){$Q^2$} \put(20,120){$Q^2F_{\pi}(Q^2)$}
\end{picture}
} \caption[]{\sf Electromagnetic pion form factor for the three
models of pion DA described in the text. Identification of the
curves follows Fig.~\ref{fig:DAmodels}. The experimental data are
from~\cite{Bebek} (blue squares) and~\cite{Fpi} (red circles). }
\label{fig:EMs}
\end{figure}

The results are shown in  Fig.~\ref{fig:EMs} in comparison with the 
experimental data \cite{Bebek,Fpi}. For this plot we have chosen 
$M^2=1.5$~GeV$^2$, $s_0^\pi=0.8$~GeV$^2$ and the factorization scale
$\mu = (1/2)(Q^2+s_0)$ as representative values; the three curves 
correspond to the models of the pion DA in Fig.~\ref{fig:DAmodels}.

The agreement is very good. Note that the oscillations at small $Q^2$ 
in model I are an artifact of the truncation of the Gegenbauer expansion.
They can be removed, e.g. by reducing $a^{\rm flat}_{12} = 0.099 \to 0.04$
which has the effect of smoothening the DA, especially 
in the central region. 

\begin{figure}[t]
\centerline{
\begin{picture}(210,140)(0,0)
\put(-5,0){\epsfxsize7.8cm\epsffile{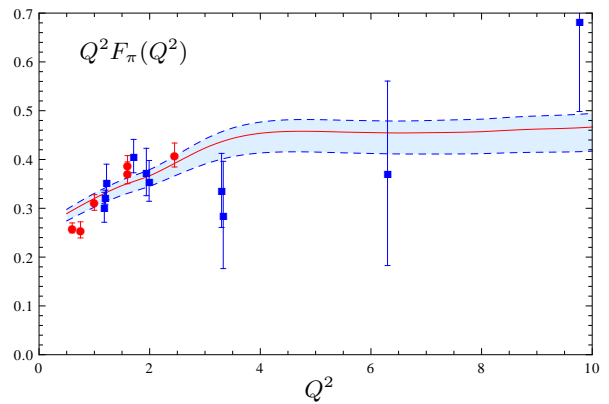}}
\put(105,-8){$Q^2$} \put(20,120){$Q^2F_{\pi}(Q^2)$}
\end{picture}
} \caption[]{\sf Electromagnetic pion form factor for model I 
(with reduced $a_{12}=0.099 \to 0.04$). The solid curve, 
upper dashed curve and lower dashed curve are calculated using 
the Borel parameter $M^2 =1.5$, $M^2=2$ and $M^2=1$~GeV$^2$,
respectively.
The experimental data are from~\cite{Bebek} (blue squares)
and~\cite{Fpi} (red circles). } \label{fig:EMcorr}
\end{figure}

It has to be mentioned that the pion electromagnetic form factor 
is much more affected by  soft contributions compared to the 
transition form factor and hence is also more model dependent. 
In particular the Borel parameter dependence is much stronger, 
see Fig.~\ref{fig:EMcorr}, where the calculations using 
$M^2=1$ and $M^2=2$~GeV$^2$ are shown by dashed curves for comparison.

The weak decay $B\to\pi \ell \nu_\ell$ has received a lot of attention 
as one of primary sources of information on the weak mixing angle $|V_{ub}|$
in the Standard Model. The differential decay width $dB/dq^2$, where 
$q^2$ is the invariant mass of the leptons, is given by the square 
of the $B\to\pi$ form factor, modulo relevant CKM angles and kinematic factors 
\begin{equation}
 \dfrac{dB}{dq^2}(B\to\pi^- e^+ \nu_e)=\dfrac{G_F^2\vert V_{ub}\vert^2}{192\pi^3m_B^3}\tau_B\,
 \lambda^{3/2}(q^2)\,\vert f_{B\pi}^+(q^2)\vert^2\,.
\end{equation}
In this equation 
$\lambda(q^2)=(m_B^2+m_\pi^2-q^2)^2-4m_B^2m_\pi^2$ and
$\tau_B$ is the mean life time of the B-meson. Below we use 
$\vert V_{ub}\vert= 3.6\cdot 10^{-3}$ to fix the overall normalization.

\begin{figure}[t]
\centerline{
\begin{picture}(210,140)(0,0)
\put(-5,0){\epsfxsize7.8cm\epsffile{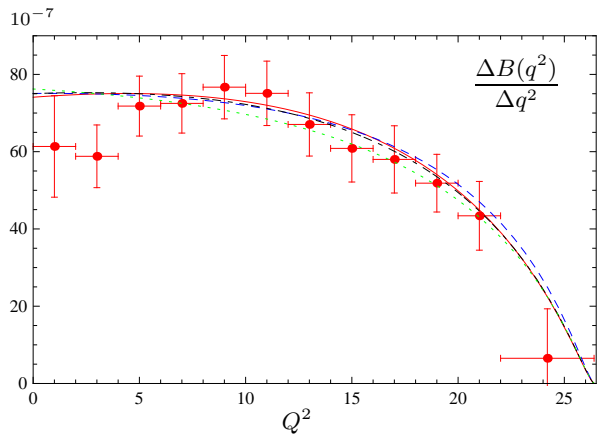}}
\put(-5,145){\scriptsize $10^{-7}$}
\put(98,-8){$Q^2$}
\put(170,120){$\dfrac{\Delta B( q^2)}{\Delta q^2} $}
\end{picture}
} \caption[]{\sf
The differential $\pi^- e^+\nu_e$ decay width. Experimental data 
are taken from~\cite{:2010zd}. A 2-parameter BCL-fit \cite{Bourrely:2008za} 
is applied to the sum rule calculation to extrapolate it to the whole $q^2$ range. 
For normalization we assume $\vert V_{ub}\vert = 3.6\cdot 10^{-3}$. 
The identification of the curves follows Fig.~\ref{fig:DAmodels}, see also text.
 } \label{fig:Bpi}
\end{figure}

LCSRs enable one to calculate the form factor up to $q^2=14$ GeV$^2$,
and a two parameter BCL \cite{Bourrely:2008za} fit is then used to
extrapolate the calculation to the whole kinematic region 
$0\leq q^2\leq 26.4$~GeV$^2$.
The latest and most advanced LCSR calculations of this form factor
\cite{Duplancic:2008ix, Ball:2004ye} include NLO corrections in leading twist and also 
for a part of the twist-3 contributions. 
Twist-4 corrections are taken into account in the leading order.
In the calculations presented in Fig.~\ref{fig:Bpi} we use the central 
values of the sum rule parameters from Ref.~\cite{Duplancic:2008ix}. 

All our models of pion DA  describe the data \cite{:2010zd} 
reasonably well and are nearly indistinguishable in the 
$q^2 < 14$ GeV$^2$ range where the direct sum rule calculation is applicable.
The calculation using ``conventional'' pion DA with $a_2=0.17$ 
and $a_4=0.06$ (at 1 GeV)~\cite{Duplancic:2008ix}
is shown by green dots for comparison.
We conclude that the $B\to\pi$ form factor is not very sensitive 
to the higher Gegenbauer moments beyond $a_2$; a low value $a_2 < 0.2$
(at 1 GeV) is preferred.

The value of the pion DA in the middle point in all our models
is close to the asymptotic value $\phi_\pi (x=1/2) = 1.5$.
This number is within the range in Eq.~(\ref{eq:middle}) and somewhat 
larger than it was assumed in the LCSR calculations of 
pion-hadron couplings  
\cite{Braun:1988qv,Belyaev:1994zk,Khodjamirian:1999hb,Aliev:1999ce,Aliev:2006xr,Aliev:2009kg}.
The larger value is in fact welcome and can reduce the well-known $\sim 30\%$ discrepancy
of the sum rule calculation \cite{Belyaev:1994zk,Khodjamirian:1999hb} of $g_{D^*D\pi }$ 
with the experiment.

\section{Summary and Conclusions}

The recent BaBar measurement \cite{BABAR} of the pion transition form factor provided 
one with the most direct evidence so far that the pion distribution amplitude 
deviates considerably from its asymptotic form. 
This result has to be considered as a success of an early QCD prediction 
\cite{Chernyak:1981zz} of a broad pion DA at a low scale, but it also 
created a lot of excitement because a significant scaling violation
at $Q^2 > 5-10$~GeV$^2$ came out unexpected.

The main lesson to be learnt from the BaBar data is that attempts 
to describe the transition form factor with one real photon entirely 
in the framework of perturbative QCD are futile; nonperturbative soft corrections 
must be taken into account. 
  
We have adopted the LCSR approach 
\cite{Balitsky:1989ry,Braun:1988qv,Chernyak:1990ag} 
which has the advantage that it is applicable to a broad class of reactions
and has been thoroughly tested. 
In this work we go beyond the existing analysis
\cite{Khodjamirian:1997tk,Schmedding:1999ap,Bakulev:2001pa,Bakulev:2002uc,Bakulev:2003cs,Agaev:2005rc,Mikhailov:2009kf}
in two aspects. First, we calculate a new, twist-six contribution
to LCSRs which proves to be sizeable. Second, we extend the
existing formalism to allow for the contributions of higher-order
Gegenbauer polynomials, which allows one to consider DAs of
arbitrary shape and also address the question of convergence of
the Gegenbauer expansion which generated some confusion.

We find that a significant rise of the scaled form factor 
$Q^2F_{\pi^0\gamma^*\gamma}(Q^2)$ in the $Q^2 = 5-20$~GeV$^2$ range 
observed by the BaBar collaboration \cite{BABAR} can be explained 
by a large value of the fourth Gegenbauer moment
$$
 a_4 > a_2
$$
in the pion DA, leading to models of the type shown in Fig.~\ref{fig:DAmodels}
which are not far from the asymptotic distribution in the central region but 
have enhancements close to the end points. Our preferred models also
include sizeable $a_6$ coefficients; these can be put to zero at the cost 
of further increasing $a_4$ which does not seem to be attractive.
The higher partial waves, $a_8$, $a_{10}$, etc. contribute only marginally in the 
BaBar $Q^2$ range, the reason being that 
contributions of the end-point regions in the pion DA  are cut off by soft effects. 

We have checked that the models of pion DA having such an inverse hierarchy,
$a_4 > a_2$ give good description of the the pion electromagnetic and weak 
decay $B\to\pi$ form factors calculated within the same LCSR approach.
This agreement is not trivial since the size of soft corrections
is very different and also the duality assumption of the contribution 
of small invariant masses is applied in different channels. 
The small value of $a_2 <  0.2$ (at 1 GeV) was actually suggested before 
from the fit to the $B\to \pi \ell\nu_\ell$ differential decay width \cite{Duplancic:2008ix},
whereas large $a_4$ does not have a noticeable effect in this case because the effective momentum 
transfer in $B$ decays is much lower.  

The main uncertainty of the LCSR calculation is due to the assumption that contributions
of low invariant masses in the dispersion relation in QCD diagrams 
are dual (i.e. coincide in integral sense) with the contribution of resonances, here 
the $\rho$, $\omega$ mesons. The accuracy of duality is difficult to quantify,
but it is usually believed to be better than 20\% on the experience of many successful 
applications. An inspection shows that our result $a_4 > a_2$ is related to a 
rather large value of the transition form factor $F_{\rho\pi\gamma^*}(Q^2)$ in the 
$Q^2\sim 2-5$ GeV$^2$ range that follows from duality. For comparison, this
form factor estimated as the integral of the spectral density below $s=1.5$~GeV$^2$
in the MR model~\cite{Musatov:1997pu} appears to be a factor 2--3 lower, which 
explains why in this model the BaBar data can be fitted by a CZ-type pion DA
with a large $a_2$ coefficient and $a_4=0$. 
Lattice calculations of the $F_{\rho\pi\gamma^*}(Q^2)$ form factor in a few GeV$^2$ range and 
improved accuracy on $a_2$ would help to discriminate between these two
possibilities. More precise experimental data in the $Q^2 =15-30$~GeV$^2$ range 
would of course be most welcome as well.   

\section*{Acknowledgements}
V.~M.~Braun is grateful to  M.~Diehl, A.~Khodjamirian, P.~Kroll
and A.~Radyushkin for numerous discussions on the subject of this
work, and to  N.~Stefanis for correspondence concerning the
results in Ref.~\cite{Schmedding:1999ap}. S.~S.~Agaev appreciates
a warm hospitality of members of the Theoretical Physics Institute
extended to him in Regensburg University, where this work has been
carried out. The work of S.~S.~Agaev was supported by the DAAD
(grant A/10/02381).


\appendix
\renewcommand{\theequation}{\Alph{section}.\arabic{equation}}

\section{Scale dependence of the pion DA}\label{App:A}

The scale dependence of the coefficients $a_n(\mu)$ in the
Gegenbauer expansion of the pion DA is determined by
Eq.~(\ref{eq:NLOevolution}).

The RG factor $E_{n}^{\mathrm{NLO}}(\mu ,\mu _{0})$
in this expression is given by
\begin{eqnarray}
&&E_{n}^{\mathrm{NLO}}(\mu ,\mu _{0})=\left[ \frac{\alpha _{\mathrm{s}}(\mu )}{%
\alpha _{\mathrm{s}}(\mu _{0})}\right] ^{\gamma _{n}^{(0)}/2\beta
_{0}} \\
&&\times \left\{ 1 +\frac{\alpha _{\mathrm{s}}(\mu )-\alpha _{\mathrm{s}}(\mu _{0})}{%
8\pi }\frac{\gamma _{n}^{(0)}}{\beta _{0}}\left( \frac{\gamma _{n}^{(1)}}{%
\gamma _{n}^{(0)}}-\frac{\beta _{1}}{\beta _{0}}\right) \right\}.
\nonumber
\end{eqnarray}
The corresponding LO RG factor  $E_{n}^{\mathrm{LO}}(\mu ,\mu
_{0})$ is obtained by keeping the first term only in the braces.

Here $\beta_0\, (\beta_1)$ and $\gamma_n^{(0)}(\gamma_n^{(1)})$ are the LO (NLO)
coefficients of the QCD $\beta$-function and the anomalous dimensions,
respectively:
\begin{eqnarray}
\mu ^{2}\frac{d\alpha _{\mathrm{s}}(\mu )}{d\mu ^{2}}&=&
\beta (\alpha _{\mathrm{s}})=-\alpha _{\mathrm{s}}
\left\{ \beta_{0}\frac{\alpha_{\mathrm{s}}}{4\pi }
+\beta_{1}\left(\frac{\alpha _{\mathrm{s}}}{4\pi}\right)^{2}+\ldots\right\}
\nonumber\\
\gamma _{n}(\alpha _{\mathrm{s}})&=&
-\frac{1}{2}\left\{ \gamma _{n}^{(0)}\frac{\alpha _{\mathrm{s}}}{4\pi }
+\gamma _{n}^{(1)}\left( \frac{\alpha _{\mathrm{s}}}{4\pi }\right) ^{2}+\ldots\right\}.
\end{eqnarray}
The first two coefficients of the beta-function are
\begin{equation}
\beta _{0}=11-\frac{2}{3}n_{f}\,,\qquad
\beta_{1}=102-\frac{38}{3}n_{f}\,,
\end{equation}%
whereas $\gamma _{n}^{(0)}$ is given by
\begin{equation}
\gamma _{n}^{(0)}=2C_{F}\left( 1-\frac{2}{(n+1)(n+2)}
+4\sum_{m=2}^{n+1}\frac{1}{m}\right).
\label{eq:anomdim0}
\end{equation}

The NLO anomalous dimensions can most easily be obtained using the
FeynCalc Mathematica package \cite{FeynCalc}. For convenience we
present explicit expressions up to $n=12$ that are used in
our calculations ($\gamma_{0}^{(1)}=0$):
\begin{eqnarray*}
 \gamma_{2}^{(1)}&=&\frac{34450}{243}-\frac{830}{81}n_{f},\\
 \gamma_{4}^{(1)}&=&\frac{662846}{3375}-\frac{31132}{2025}n_{f}, \\
 \gamma_{6}^{(1)} &=&\frac{718751707}{3087000}
        -\frac{3745727}{198450}n_{f}, \\
\gamma_{8}^{(1)} &=&\frac{293323294583}{1125211500}
-\frac{19247947}{893025}n_{f}, \\
\gamma_{10}^{(1)} &=&\frac{212204133652373}{748828253250}
-\frac{512808781}{21611205}n_{f}, \\
\gamma_{12}^{(1)} &=&\frac{995653107122188087}{3290351344780500}
-\frac{93360116539}{3652293645}n_{f}.
\end{eqnarray*}
The off-diagonal mixing coefficients $d_n^k$ in Eq.~(\ref{eq:NLOevolution})
are given by the following expression:
\begin{eqnarray}
d_{n}^{k}(\mu ,\mu _{0})&=&\frac{M_{n}^{k}}
{\gamma_{n}^{(0)}-\gamma_{k}^{(0)}-2\beta_{0}}
\nonumber \\
&&\hspace*{-0.7cm}{}\times \left\{ 1
-\left[ \frac{\alpha _{\mathrm{s}}(\mu)}
{\alpha _{\mathrm{s}}(\mu _{0})}\right] ^{[\gamma_{n}^{(0)}-\gamma _{k}^{(0)}-2\beta _{0}]/2\beta _{0}}\right\}.
\label{eq:a.2}
\end{eqnarray}
The matrix $M_{n}^{k}$ is defined as
\begin{eqnarray}
M_{n}^{k}&=&\frac{(k+1)(k+2)(2n+3)}{(n+1)(n+2)}\left[\gamma_{n}^{(0)}-\gamma _{k}^{(0)}\right]
\nonumber\\&&\hspace*{-1cm}
\times\left\{ \frac{8C_{F}A_{n}^{k}-\gamma _{k}^{(0)}-2\beta _{0}}{(n-k)(n+k+3)}
+4C_F\frac{A_{n}^{k}-\psi (n+2)+\psi (1)}{(k+1)(k+2)}\right\}
\nonumber\\
\label{eq:a.3}
\end{eqnarray}
where
\begin{eqnarray}
A_{n}^{k} &=&\psi \Big( \frac{n+k+4}{2}\Big) -\psi \Big( \frac{n-k}{2}\Big)
\nonumber \\
&&{}+2\psi(n-k)-\psi(n+2)-\psi(1)\,.
\label{eq:a.4}
\end{eqnarray}
{}For convenience, we have collected numerical values of the
coefficients $M_{n}^{k}$ for $n\le 12$ in Table~\ref{tab:mnk}.

\begin{table*}[t]
\renewcommand{\arraystretch}{1.2}
\begin{center}
\begin{tabular}{|c|c|c|c|c|c|c|} \hline
$M_n^{k}$& $k=0$ & $k=2$ & $k=4$ & $k=6$ & $k=8$ & $k=10$ \\ \hline
$n=0$ & $0$ & & &  &  &    \\
\hline $n=2$ & $-11.23+1.73n_{f}$ & $0$ & &
& &
\\ \hline
$n=4$ & $-1.41+0.56n_{f}$ & $-22.02+1.65n_{f}$ & $0$ &  &  &
\\ \hline
$n=6$
& $\phantom{-}0.03+0.26n_{f}$ & $-7.76+0.82n_{f}$ & $-22.77+1.39n_{f}$ &
$0$ & & \\
\hline
$n=8$
& $\phantom{-}0.29+0.14n_{f}$ & $-3.34+0.48n_{f}$ &
$-10.34+0.84n_{f}$ & $-21.72+1.18n_{f}$ & $0$ &    \\
\hline
$n=10$
& $\phantom{-}0.31+0.09n_{f}$ & $-1.58+0.30n_{f}$ & $-5.46+0.55n_{f}$ &
$-11.3+0.79n_{f}$ & $-20.35+1.02n_{f}$ & $0$  \\
\hline
$n=12$ &
$\phantom{-}0.28+0.06n_{f}$ & $-0.78+0.21n_{f}$ & $-3.13+0.38n_{f}$ &
$-6.64+0.56n_{f}$ & $-11.54+0.73n_{f}$ & $-19.0+0.9n_{f}$ \\
\hline
\end{tabular}
\end{center}
\caption[]{\sf The mixing matrix $M_n^{k}$ (\ref{eq:a.3}).}
\label{tab:mnk}
\renewcommand{\arraystretch}{1.0}
\end{table*}

\section{The NLO spectral density}\label{App:B}

The coefficients $G_n^{k}$ and $H_n^{k}$ in the expansion of the
NLO perturbative spectral density (\ref{eq:ABOP}) are collected in
Tabs.~\ref{tab:Gnk} and \ref{tab:Hnk}, respectively. Our results
for $G_n^k$ agree with Ref.~\cite{Mikhailov:2009kf} (except for
$G_0^0$ and $G_4^0$) noting an overall sign difference in
definition of $G_n^k$ , whereas for $H_n^k$ the difference is that
the expansion in Eq.~(\ref{eq:ABOP}) also involves contributions
with odd $k=2\ell+1$.

\begin{table*}[tb]
\renewcommand{\arraystretch}{1.5}
\begin{center}
{\large
\begin{tabular}{|c|c|c|c|c|c|c|c|} \hline
{\normalsize $G_n^{k}$}&
{\normalsize $k=0$}&
{\normalsize $k=2$}&
{\normalsize $k=4$}&
{\normalsize $k=6$}&
{\normalsize $k=8$}&
{\normalsize $k=10$}&
{\normalsize $k=12$}
\\ \hline
{\normalsize $n=0$}
&{\normalsize $-1$} & & &  &  &  &
\\ \hline
{\normalsize $n=2$} & $\frac{3}{2}$ &
$-\frac{35}{12}$ & & & & &
\\ \hline
{\normalsize $n=4$}
 & $\frac{3}{4}$ & $\frac{161}{72}$ & $-\frac{203}{45}$ &  &  &  &
\\ \hline
{\normalsize $n=6$}
 & $\frac{83}{180}$ & $\frac{49}{40}$ & $\frac{781}{300}$ &
$-\frac{29531}{5040}$ & & &
\\ \hline
{\normalsize $n=8$}
 & $\frac{177}{560}$ & $\frac{4}{5}$ &
$\frac{6259}{4200}$ & $\frac{4437}{1568}$ &
$-\frac{177133}{25200}$ & &
\\ \hline
{\normalsize $n=10$}
 & $\frac{487}{2100}$ & $\frac{6181}{10800}$ &
$\frac{7601}{7560}$ & $\frac{7823}{4704}$ &
$\frac{338561}{113400}$ & $-\frac{1676701}{207900}$ &
\\ \hline
{\normalsize $n=12$}
 & $\frac{74141}{415800}$ & $\frac{17167}{39600}$ & $\frac{697}{945}$ & $\frac{177799}{155232}$ &
  $\frac{2227921}{1247400}$ & $\frac{5672237}{1829520}$ & $-\frac{30946717}{3439800}$ \\
\hline
\end{tabular}
}
\end{center}
\caption[]{\sf Numerical values of the coefficients $G_n^{k}$ (\ref{eq:Gkn}).} \label{tab:Gnk}
\renewcommand{\arraystretch}{1.0}
\end{table*}

\begin{table*}[tb]
\renewcommand{\arraystretch}{1.5}
\begin{center}
{\large
\begin{tabular}{|c|c|c|c|c|c|c|c|c|c|c|c|c|c|} \hline
{\normalsize $H_n^{k}$}&
{\normalsize $k=0$}&
{\normalsize $k=1$}&
{\normalsize $k=2$}&
{\normalsize $k=3$}&
{\normalsize $k=4$}&
{\normalsize $k=5$}&
{\normalsize $k=6$}&
{\normalsize $k=7$}&
{\normalsize $k=8$}&
{\normalsize $k=9$}&
{\normalsize $k=10$}&
{\normalsize $k=11$}&
{\normalsize $k=12$}
\\\hline
{\normalsize $n=0$}& $\frac{3}{2}$ & & & & & & & & & & & &
\\ \hline
{\normalsize $n=2$} &
$\frac{3}{2}$ & $-\frac{5}{2}$ &$\frac{25}{12}$ & & & &  & &  & & & &\\
\hline
{\normalsize $n=4$} &
$\frac{3}{2}$ & $-\frac{5}{4}$ & $\frac{7}{12}$
&$-\frac{9}{4}$ &$\frac{49}{20}$ & & & & & & & &
\\ \hline
{\normalsize $n=6$} &
$\frac{3}{2}$ & $-\frac{31}{30}$ & $\frac{7}{12}$ &
$-\frac{19}{20}$ &$\frac{11}{30}$ &$-\frac{13}{6}$
&$\frac{761}{280}$ & & & & & &
\\ \hline
{\normalsize $n=8$} &
$\frac{3}{2}$ & $-\frac{20}{21}$ &
$\frac{7}{12}$ & $-\frac{99}{140}$ & $\frac{11}{30}$
&$-\frac{143}{168}$ &$\frac{15}{56}$ &$-\frac{17}{8}$
&$\frac{7381}{2520}$ & & & &
\\  \hline
{\normalsize $n=10$} &
$\frac{3}{2}$ & $-\frac{115}{126}$ &
$\frac{7}{12}$ & $-\frac{171}{280}$ & $\frac{11}{30}$ &
$-\frac{377}{630}$ &$\frac{15}{56}$ &$-\frac{289}{360}$
&$\frac{19}{90}$ &$-\frac{21}{10}$ &$\frac{86021}{27720}$ & &
\\ \hline
{\normalsize $n=12$} &
$\frac{3}{2}$ & $-\frac{235}{264}$ &
$\frac{7}{12}$ & $-\frac{101}{180}$ & $\frac{11}{30}$ &
$-\frac{52}{105}$ & $\frac{15}{56}$
&$-\frac{2159}{3960}$ &$\frac{19}{90}$ &$-\frac{511}{660}$ &$\frac{23}{132}$ &$-\frac{25}{12}$ &$\frac{1171733}{360360}$ \\
\hline
\end{tabular}
}
\end{center}
\caption[]{\sf Numerical values of the coefficients $H_n^{k}$ (\ref{eq:Hkn}).} \label{tab:Hnk}
\renewcommand{\arraystretch}{1.0}
\end{table*}




\end{document}